\documentclass[review,12pt]{elsarticle}

\usepackage{amssymb}
\usepackage{amsthm}
\usepackage{framed} 
\usepackage{amsmath,color}
\usepackage{mathrsfs}
\usepackage{graphicx}
\usepackage{epstopdf}
\usepackage{float}
\usepackage{bm}
\usepackage{bbm}
\usepackage{mathrsfs}
\usepackage{cleveref}
\usepackage{caption}
\usepackage{subcaption}
\usepackage{soul}
\usepackage{accents}
\usepackage{color,soul} 
\usepackage{color} 
\usepackage{nomencl} 
\makenomenclature 
\biboptions{sort&compress}
\soulregister\citep7 
\soulregister\citet7 
\soulregister\citealp7 
\newsavebox{\measurebox} 
\usepackage{titlesec} 
\usepackage{tabu} 
\usepackage{longtable}
\usepackage{tikz}
\usepackage{multirow}
\tikzset{>=latex}
\usepackage{pstricks}
\usepackage{listings}
\usepackage[margin=3cm]{geometry}
\journal{Composites Part B: Engineering}

\makeatletter
\def\@author#1{\g@addto@macro\elsauthors{\normalsize%
    \def\baselinestretch{1}%
    \upshape\authorsep#1\unskip\textsuperscript{%
      \ifx\@fnmark\@empty\else\unskip\sep\@fnmark\let\sep=,\fi
      \ifx\@corref\@empty\else\unskip\sep\@corref\let\sep=,\fi
      }%
    \def\authorsep{\unskip,\space}%
    \global\let\@fnmark\@empty
    \global\let\@corref\@empty  
    \global\let\sep\@empty}%
    \@eadauthor={#1}
}
\makeatother

\titleformat{\paragraph}
{\normalfont\normalsize\itshape}{\theparagraph}{1em}{}
\titlespacing*{\paragraph}
{0pt}{3.25ex plus 1ex minus .2ex}{1.5ex plus .2ex}

\begin{document}

\begin{frontmatter}



\title{Phase field modelling of crack propagation in functionally graded materials}

\author{Hirshikesh\fnref{IITM}}

\author{Sundararajan Natarajan\fnref{IITM}}

\author{Ratna K. Annabattula\fnref{IITM}}

\author{Emilio Mart\'{\i}nez-Pa\~neda\corref{cor1}\fnref{Cam}}
\ead{mail@empaneda.com}

\address[IITM]{Department of Mechanical Engineering, Indian Institute of Technology Madras, Chennai - 600036, India}

\address[Cam]{Department of Engineering, Cambridge University, CB2 1PZ Cambridge, UK}

\cortext[cor1]{Corresponding author.}

\begin{abstract}
We present a phase field formulation for fracture in functionally graded materials (FGMs). The model builds upon homogenization theory and accounts for the spatial variation of elastic and fracture properties. Several paradigmatic case studies are addressed to demonstrate the potential of the proposed modelling framework. Specifically, we (i) gain insight into the crack growth resistance of FGMs by conducting numerical experiments over a wide range of material gradation profiles and orientations, (ii) accurately reproduce the crack trajectories observed in graded photodegradable copolymers and glass-filled epoxy FGMs, (iii) benchmark our predictions with results from alternative numerical methodologies, and (iv) model complex crack paths and failure in three dimensional functionally graded solids. The suitability of phase field fracture methods in capturing the crack deflections intrinsic to crack tip mode-mixity due to material gradients is demonstrated. Material gradient profiles that prevent unstable fracture and enhance crack growth resistance are identified: this provides the foundation for the design of fracture resistant FGMs. The finite element code developed can be downloaded from www.empaneda.com/codes.
\end{abstract}


\end{frontmatter}



\section{Introduction}
\label{Sec:Introduction}
Functionally graded materials (FGMs) are multifunctional composites with spatially varying volume fractions of constituent materials. The resulting graded macroproperties enable designers to tailor the microstructure to specific operating conditions, while minimizing problems associated with discrete material interfaces. Potential advantages inherent to the use of FGMs include, among others, enhancing the thermal stress resistance \cite{Aboudi1994,Pindera1998}, reducing residual stresses \cite{Lee1994}, and suppressing stress discontinuities at bimaterial interface structures \cite{Ramaswamy1997,Tilbrook2006}. As a consequence, FGMs have found a wide range of commercial applications including cutting tools, biomedical devices, optical fibers and wear resistant coatings \cite{Pindera1997,Uemura2003}. However, FGMs commonly exhibit brittle fracture, as ceramic materials are one of the most frequently used constituents. Hence, fracture resistance often constitutes the primary design criterion, and a special branch of fracture mechanics has been devoted to investigate failure in this class of materials \cite{Erdogan1995}.\\

Understanding crack initiation and subsequent growth in functionally graded components is hindered by the influence of material in-homogeneity on the crack propagation path. For example, a crack will deviate from the self-similar crack propagation direction in a functionally graded specimen subjected to pure mode I loading if the material gradient is not aligned with the crack. This crack tip mode mixity induced by the spatial variation of material properties leads to complex crack trajectories that are difficult to capture with conventional numerical methods. The vast majority of works have tried to address this challenge by using discrete numerical methods; efforts include the use of remeshing algorithms \cite{Kim2004}, extended finite element methods (X-FEM) \cite{Comi2007,Bayesteh2013,IJMMD2015}, scaled boundary finite element formulations \cite{Ooi2015}, and cohesive zone models \cite{Jin2003}. However, these techniques are limited in tracking the crack path topology, particularly in 3D problems. Variational approaches based on energy minimization constitute a promising tool to overcome this limitation and could, therefore, be particularly useful in modelling crack advance in FGMs \cite{Francfort1998,Doan2016,VanDo2017}. Specifically, the phase field method has proven to be a compelling technique in modelling brittle fracture \citep{McAuliffe2016,Bleyer2018}, ductile damage \cite{Borden2016,Miehe2016b}, fiber cracking and composites delamination \cite{Reinoso2017a,Carollo2017}, and hydrogen embrittlement \citep{Duda2018,CMAME2018}, among other phenomena. We aim at extending this success by presenting a phase field formulation for fracture in compositionally graded materials.

\section{A phase field fracture formulation for FGMs}
\label{Sec:NumModel}

\subsection{Governing balance equations}

Consider a functionally graded solid occupying the domain $\Omega$ with a discontinuous surface $\Gamma$. The elastic and fracture properties of the solid vary gradually in space such that the energy contributions in the deformation-fracture problem can be characterized by a strain energy density $\psi \left( \bm{x} \right)$ and a critical energy release rate $G_c \left( \bm{x} \right)$. The discrete crack is approximated through an auxiliary field variable $\phi$, so-called phase field parameter, which varies between $\phi=0$, intact material, and $\phi=1$, completely damaged. The size of the regularized crack surface is governed by the choice of $\ell$, the phase field model-inherent length scale. As shown by $\Gamma$-convergence \cite{Bellettini1994}, a regularized crack density functional $\Gamma_\ell \left( \ell, \phi \right)$ can be defined that converges to the functional of the discrete crack as $\ell \to 0$. Hence, the fracture energy due to the creation of a crack can be approximated as
\begin{equation}
\int_\Gamma G_c \left( \bm{x} \right) \, \text{d} \Gamma  \approx \int_\Omega G_c \left( \bm{x} \right) \Gamma_{\ell} \left( \ell, \phi \right) \, \text{d} \Omega = \int_\Omega G_c \left( \bm{x} \right) \left( \frac{1}{2\ell}\phi^2 + \frac{\ell}{2} |\nabla \phi|^2 \right) \text{d} \Omega
\label{eq:surfaceenergy}
\end{equation}

The approximated crack surface energy (\ref{eq:surfaceenergy}) can be added to the bulk energy to form the total potential energy of the functionally graded solid $\Psi$ as
\begin{equation}\label{eq:Psi2}
    \Psi = \int_\Omega \left(  (1- \phi)^2 \, \psi (\bm{\varepsilon}, \bm{x}) + G_c \left( \bm{x} \right) \left(  \frac{1}{2\ell}\phi^2 + \frac{\ell}{2} |\nabla \phi|^2  \right)  \right) \text{d} \Omega
\end{equation}

\noindent where the term $(1- \phi)^2$ describes the degradation of the stored energy with evolving damage. The strain energy density for the undamaged solid is given in terms of the strain field $\bm{\varepsilon}$ and the \emph{spatially-varying} linear elastic stiffness matrix $\bm{C} \left( \bm{x} \right)$ as
\begin{equation}
    \psi (\bm{\varepsilon}, \bm{x}) = \frac{1}{2} \bm{\varepsilon}^T : \bm{C} \left( \bm{x} \right) : \bm{\varepsilon} 
\end{equation}

\noindent with the strain tensor being related to the displacement field in the usual manner: $\bm{\varepsilon}= \text{sym} \nabla \bm{u}$. Upon taking the first variation of (\ref{eq:Psi2}) and applying Gauss theorem, the following coupled field equations are obtained for any arbitrary value of the kinematic variables $\delta \bm{u}$ and $\delta \phi$,
\begin{align}\label{eqn:strongForm}
(1-\phi)^2 \, \, \nabla \cdot \boldsymbol{\sigma}  &= \boldsymbol{0}   \hspace{3mm} \rm{in}  \hspace{3mm} \Omega \\ \nonumber
G_{c} \left( \bm{x} \right)  \left( \dfrac{1}{\ell} \phi - \ell \Delta \phi \right) - 2(1-\phi) \, \psi \left(\bm{\varepsilon}, \bm{x} \right) &= 0 \hspace{3mm} \rm{in}  \hspace{3mm} \Omega
\end{align}

\noindent Here, $\bm{\sigma}$ denotes the Cauchy stress tensor.

\subsection{Homogenization scheme}
The stiffness and fracture resistance dependence on $\bm{x}$ inherent to FGMs can be inferred from the spatial variation of the volume fractions of constituent materials via homogenization. Consider a functionally graded beam with thickness $h$ and material gradation along a $y$-axis that gradually changes from 100\% volume fraction of compound 1 to 100\% volume fraction of compound 2. The volume fraction of material 1, $V_1$, reads,
\begin{equation}\label{eq:VolumeFraction}
V_1 = \left(\frac{1}{2} + \frac{y}{h} \right)^k
\end{equation}

\noindent where $k$ is the material gradient index or volume fraction exponent. From the volume fraction, the local effective properties can be obtained by means of a Mori-Tanaka homogenization scheme. Hence, the effective bulk modulus $K_e$ and shear modulus $\mu_e$ read
\begin{equation}
\frac{K_e-K_1}{K_2 - K_1} = \frac{V_2}{1+3 V_1  \left(K_2 - K_1 \right) / \left( 3 K_1 + 4 \mu_1 \right)}
\end{equation}
\begin{equation}
\frac{\mu_e-\mu_1}{\mu_2-\mu_1} = \frac{V_2}{1+V_1 \left(\mu_2 - \mu_1 \right)/\left(\mu_1 + \mu_1 \left( 9 K_1 + 8 \mu_1 \right) / \left( 6 \left( K_1 + 2 \mu_1 \right) \right) \right)}
\end{equation}

\noindent And from $K_e$ and $\mu_e$, one can readily compute the effective Young's modulus $E_e$ and Poisson's ratio $\nu_e$ using the standard relations.\\

The fracture response in the phase field model is governed by the critical energy release rate $G_c$ and the length scale parameter $\ell$. While the latter is often seen as a regularizing parameter, its value influences the critical stress at which damage initiates, $\sigma_f$. For example, the analytical homogeneous solution to a one-dimensional quasi-static problem renders \cite{CMAME2018}
\begin{equation}
    \ell = \frac{27 E G_c}{256 \sigma_f}
\end{equation}

\noindent such that $\ell$ depends on $G_c \left( \bm{x} \right)$, $E \left( \bm{x} \right)$ and $\sigma_f \left( \bm{x} \right)$, which are spatially varying material properties. However, in a linear elastic, ideally brittle solid one can relate the remote load at which the failure stress is attained $K_{Ic}$ with Griffith's critical energy release rate; for example, assuming plane strain conditions:
\begin{equation}\label{Eq:GcK}
    G_c=\frac{\left( 1 - \nu^2 \right) K_{Ic}^2}{E}
\end{equation}

\noindent Given the proportional relation between the fracture toughness and the failure stress, $K_{Ic} \sim \sigma_f \sqrt{\pi a}$, this necessarily implies that the spatial dependence of $\ell$ can be neglected. The spatially varying fracture resistance of FGMs is therefore captured by $G_c \left( \bm{x} \right)$. The critical energy release rate variation can be computed as a function of the volume fraction of the constituents by using the rule of mixtures,
\begin{equation}
{G_c}_e = {G_c}_1 V_1 + {G_c}_2 V_2
\end{equation}

\noindent where ${G_c}_1$ and ${G_c}_2$ respectively denote the critical energy release rate of material 1 and material 2.

\subsection{Finite element implementation}

The finite element method is used to solved the coupled system of equations (\ref{eqn:strongForm}). The variation of elastic and fracture properties is defined at each integration point, rendering a so-called \emph{graded} finite element implementation . The phase field method can also be used to capture crack growth in layered FGMs, where crack tip mode-mixity depends on the modulus mismatch between the layers \cite{Fleck1991}. However, a smooth material property variation is best captured by implementing the material gradient at the integration point level \cite{Materials2019}. To ensure irreversibility of the phase field parameter we follow Ambati et al. \cite{Ambati2015} and adopt a hybrid approach. Thus, we reformulate Eq. (\ref{eqn:strongForm}b) as
\begin{equation}
    G_{c} (\bm{x}) \left(\dfrac{1}{\ell} \phi - \ell \Delta \phi \right) - 2(1-\phi) H^+(\bm{\varepsilon}, \bm{x}) = 0
\end{equation}

\noindent where $H^+$ is the so-called history variable field, which is defined as
\begin{equation}
    H^+= \underset{\tau \in [0, t]}{\text{max}} \psi_0^+(\bm{\varepsilon}(\bm{x}, \tau))
\end{equation}

\noindent with
\begin{equation}
   \psi_0^{\pm}(\bm{\varepsilon}, \bm{x}) = \frac{1}{2} \lambda \langle \text{tr} (\bm{\varepsilon}) \rangle^{2}_{\pm}  + \mu \text{tr} (\bm{\varepsilon}^{2}_{\pm}) 
\end{equation}

\noindent Here, $\langle a \rangle_{\pm}= \frac{1}{2}(a \pm |a|)$ and $\boldsymbol{\varepsilon}_{\pm}= \sum\limits^3_{I = 1} \langle \varepsilon_I \rangle_{\pm} \bm{n}_I \otimes \bm{n}_I$, where $\varepsilon_I$ and $\bm{n}_I$ are the principal strains and the principal strain directions, respectively. The resulting weak form can be obtained by considering the dimensional trial ($\mathscr{U}, \mathscr{P})$ and test spaces ($\mathscr{V},\mathscr{Q}$). Let $\mathscr{W}(\Omega)$ include the linear displacement field and phase field variable:
\begin{subequations}
\begin{align}
(\mathscr{U},\mathscr{V}) &=
\left\{ (\bm{u},\bm{v}) \in [ C^0(\Omega)]^d : (\bm{u},\bm{v}) \in
[ \mathcal{W}(\Omega)]^d \subseteq [ H^{1}(\Omega)]^d \right\}  \\
(\mathscr{P},\mathscr{Q}) &= \left\{ (\phi,q) \in
[ C^0(\Omega) ]^d : (\phi, q) \in [ \mathcal{W}(\Omega)]^d \subseteq
[ H^{1}(\Omega) ]^d
\right\}
\end{align}
\end{subequations}

The system of equations can be readily obtained upon applying the standard Bubnov-Galerkin procedure. In the absence of remote tractions and body forces, one can find $\bm{u} \in \mathscr{U} \, \& \, \phi \in \mathscr{P}$, for all $\bm{v} \in \mathscr{V} \, \& \, q \in \mathscr{Q}$, by solving
\begin{equation}
    \int_{\Omega} \left\{ (1-\phi)^2 \bm{\sigma}(\bm{u}):\bm{\varepsilon}(\bm{v}) \right\} \text{d} \Omega  = 0
\end{equation}
\begin{equation}
    \int_{\Omega} \left\{ \nabla q \cdot \nabla \phi \, G_c \, \ell  + q \left( \frac{G_c}{\ell} + 2 H^+ \right) \phi - 2H^+ q   \right\} \, \text{d} \Omega = 0
\end{equation}

The system is solved by means of a staggered approach using the open source finite element package FEniCS \cite{Alnaes2007,Hirshikesh2018a,Hirshikesh2019}. The script developed can be downloaded from www.empaneda.com/codes, including comprehensive documentation and verification examples.
 
\section{Results}
\label{Sec:Results}

We demonstrate the capabilities of the present framework by addressing five boundary value problems. First, the influence of the material gradient is assessed in Section \ref{Sec:CaseStudy1} by modelling crack propagation in an alumina-zirconia FGM under different gradation profiles and orientations. Insight is also gained by comparing with experiments conducted on FGM specimens. Numerical predictions show a remarkable degree of agreement with experiments on graded photodegradable copolymers (Section \ref{Sec:CaseStudy2}), and graded glass-filled epoxy composites (Section \ref{Sec:CaseStudy3}); spanning a wide range of FGMs. The potential of the present modelling approach is showcased by comparing with other numerical methods in the prediction of complex crack propagation paths in functionally graded specimens, Section \ref{Sec:CaseStudy4}, as well as addressing for the first time crack growth in three-dimensional problems with different gradient profiles, Section \ref{Sec:CaseStudy5}.

\subsection{Cracking of an alumina/zirconia plate}
\label{Sec:CaseStudy1}

First, fracture in an alumina/zirconia functionally graded plate is investigated. The functionally graded solid is built by varying the volume fractions of alumina Al$_2$O$_3$ and zirconia ZrO$_2$, as given by the homogenization scheme outlined in Section \ref{Sec:NumModel}. The elastic and fracture properties of Al$_2$O$_3$ and ZrO$_2$ are given in Table \ref{Tab:AlO3ZrO2} (see, e.g., \cite{Hadraba2004}).

\begin{table}[H]
\centering
\caption{Material properties of alumina and zirconia.}
\begin{tabular}{lcc}
\hline
                                                  & Al$_2$O$_3$ & ZrO$_2$ \\ \hline
Young's modulus, E (GPa)                         & 380.0          & 210.0        \\ 
Poisson's ratio, $\nu$                            & 0.26           & 0.31         \\ 
Fracture toughness, $K_{\rm{IC}}$ (MPa$\sqrt m$) & 5.2            & 9.6          \\ \hline
\end{tabular}\label{Tab:AlO3ZrO2}
\end{table}

\begin{figure}[H]
        \centering
        \begin{subfigure}[h]{0.49\textwidth}
                \centering
                \includegraphics[scale=0.5]{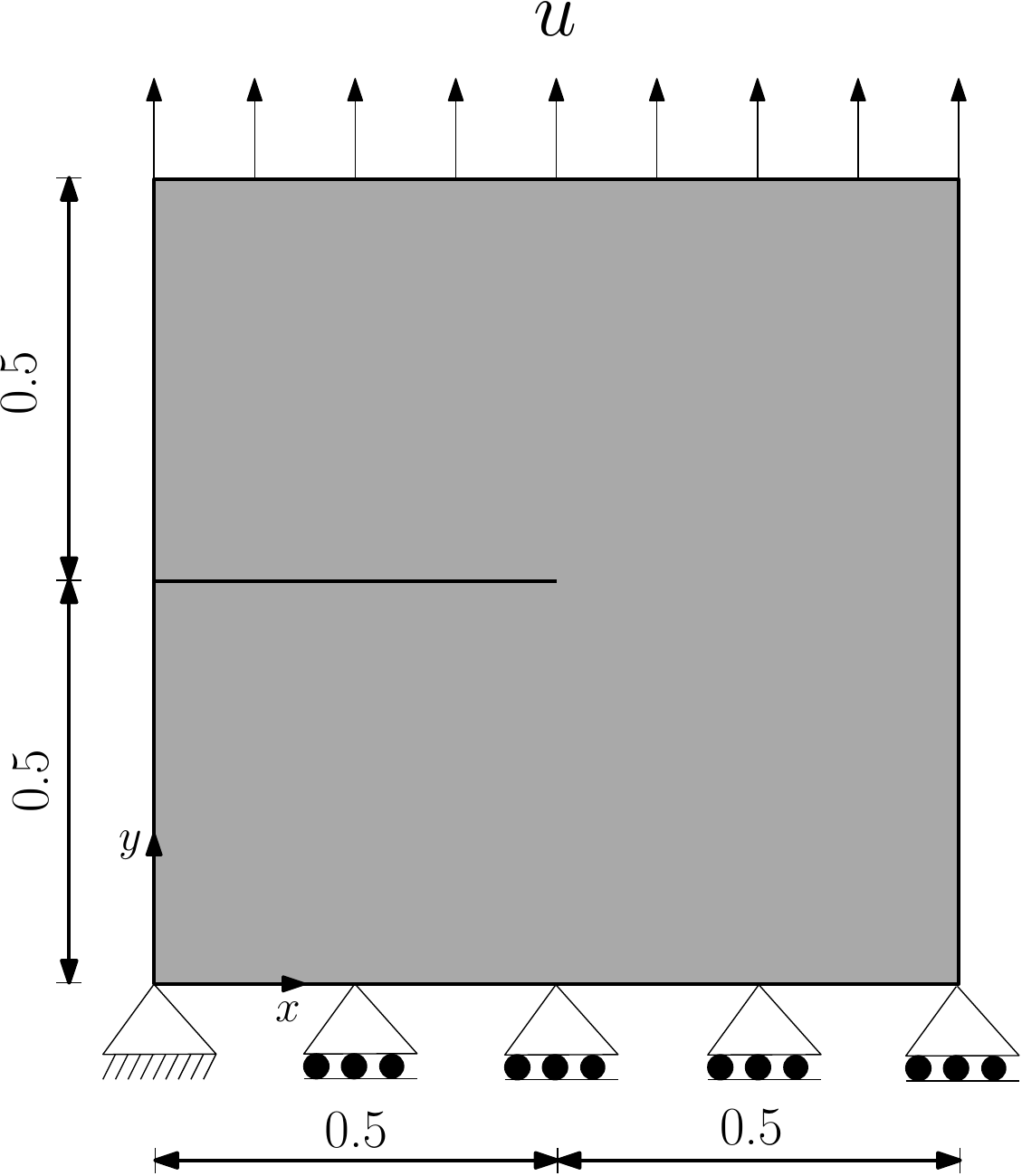}
                \caption{}
                \label{fig:domain_edgecrack}
        \end{subfigure}        
        \begin{subfigure}[h]{0.49\textwidth}
                \centering
                \includegraphics[scale=0.6]{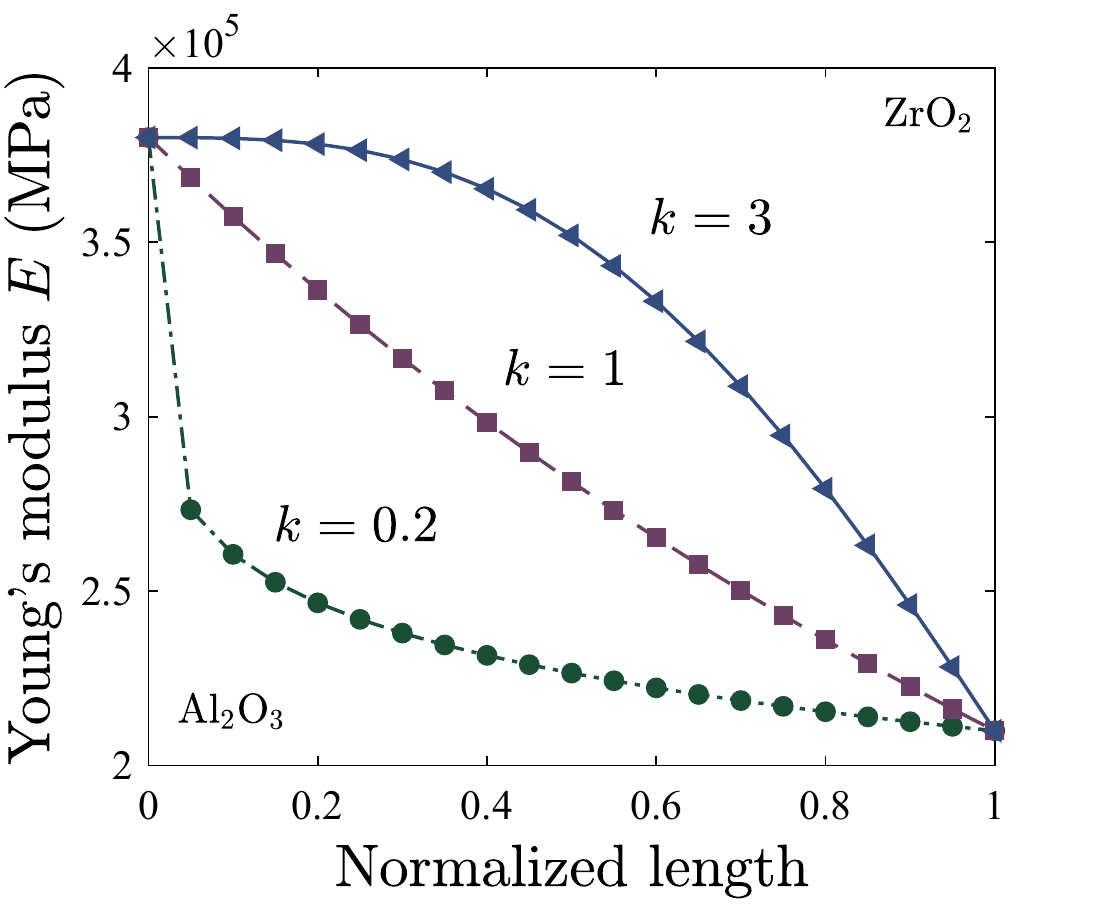}
                \caption{}
                \label{fig:Evariation}
        \end{subfigure}\\
        \vspace{10pt}
        \begin{subfigure}[h]{0.49\textwidth}
                \centering
                \includegraphics[scale=0.6]{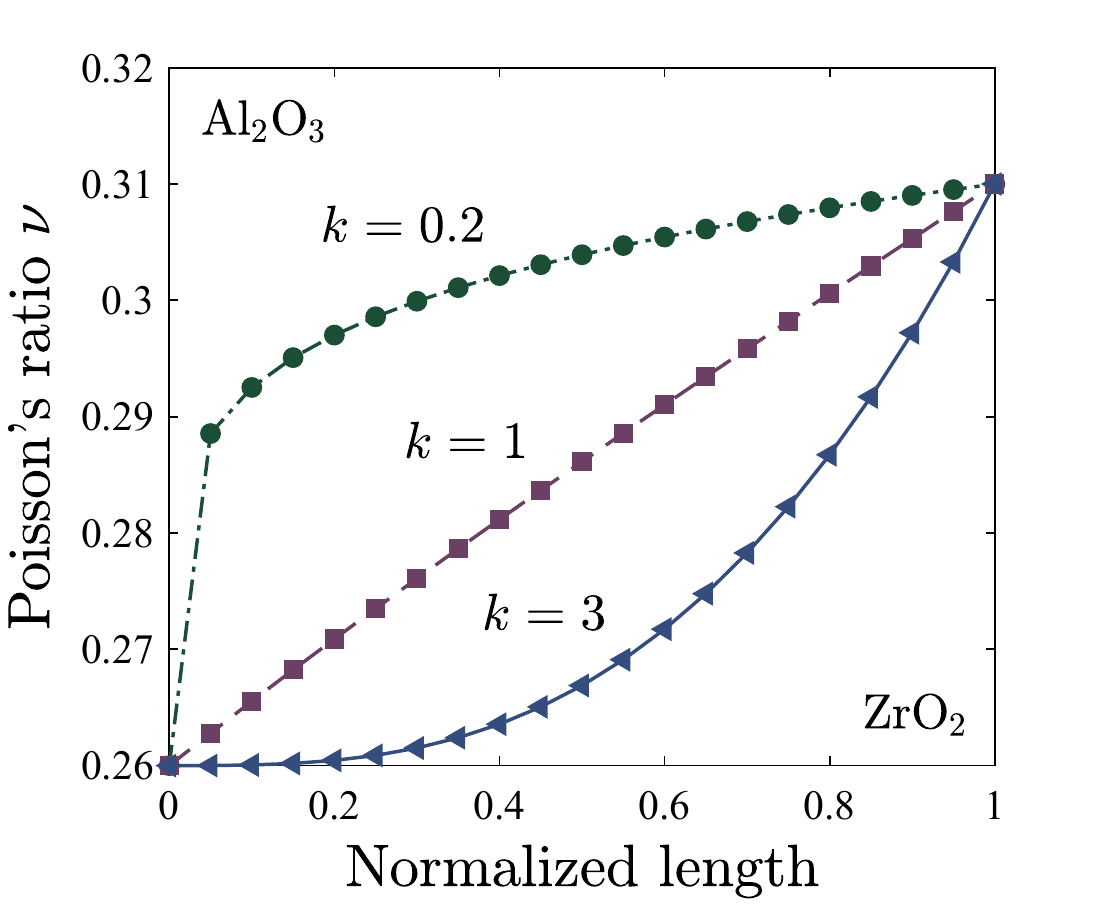}
                \caption{}
                \label{fig:Nuvariation}
        \end{subfigure}        
        \begin{subfigure}[h]{0.49\textwidth}
                \centering
                \includegraphics[scale=0.6]{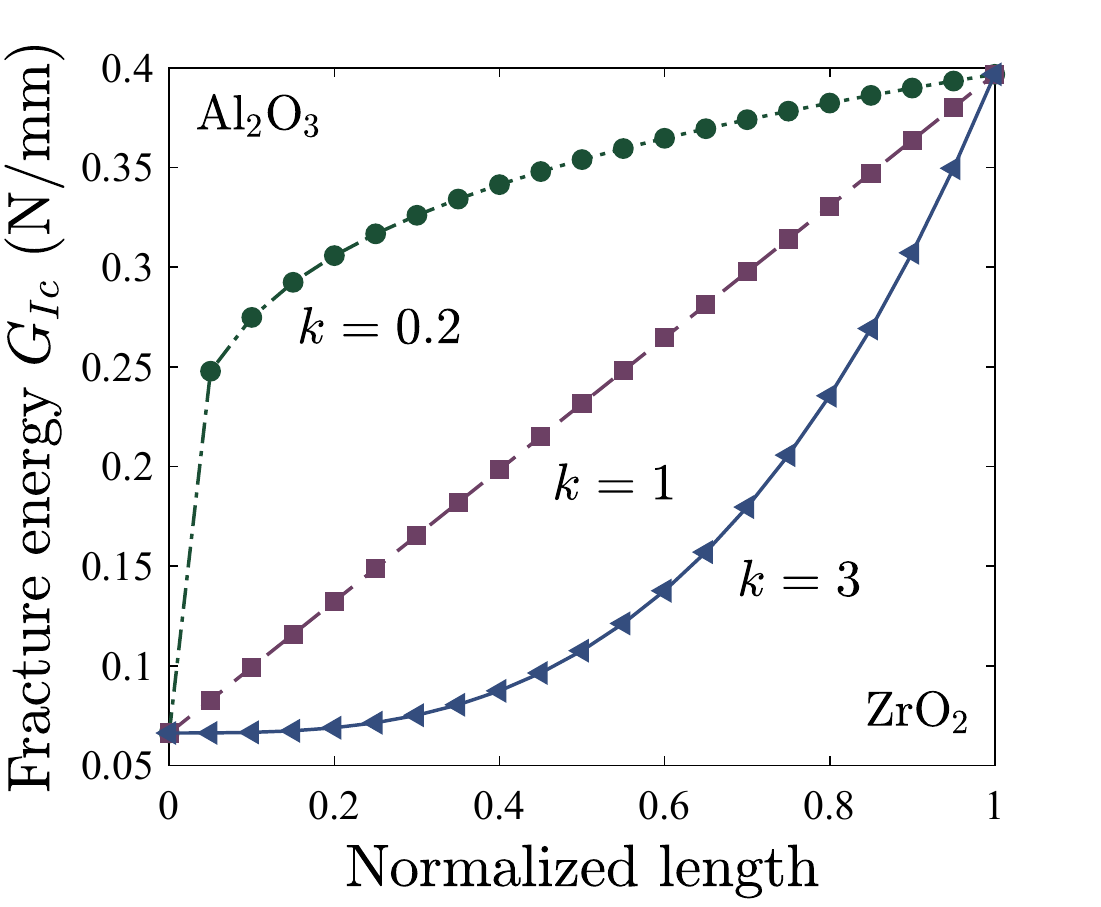}
                \caption{}
                \label{fig:Gcvariation}
        \end{subfigure}
        \caption{Al$_2$O$_3$/ZrO$_2$ graded plate, (a) geometry and boundary conditions, with dimensions given in mm, and material property variation for (b) Young's modulus $E$, (c) Poisson's ratio $\nu$ and (d) critical energy release rate $G_c$. Normalized lengths 0 and 1 respectively denote 100\% Al$_2$O$_3$ and 100\% ZrO$_2$ contents.}\label{fig:Case1matproperties}
\end{figure}

The FGM plate has a mode I edge crack and is subjected to remote uniaxial displacement, see Fig. \ref{fig:domain_edgecrack}. Plane strain conditions are assumed and the variation in space of the critical energy release rate $G_c$ is obtained by using the rule of mixtures and Eq. (\ref{Eq:GcK}). Different volume fraction profiles are considered, as given by selected values of the volume fraction exponent $k=0.2,1,3$. The variation of $E$, $\nu$ and $G_c$ along the material gradation direction is shown in Fig. \ref{fig:Case1matproperties} as a function of the material gradient index $k$. The phase field model inherent length scale equals $\ell$ = 0.003 mm; throughout the work, the element size is chosen to be at least 5 times smaller than $\ell$, so as to resolve the fracture process zone. The model contains a total of approximately 300000 degrees of freedom.

\begin{figure}[H]
\makebox[\linewidth][c]{%
        \begin{subfigure}[b]{0.55\textwidth}
                \centering
                \includegraphics[scale=0.7]{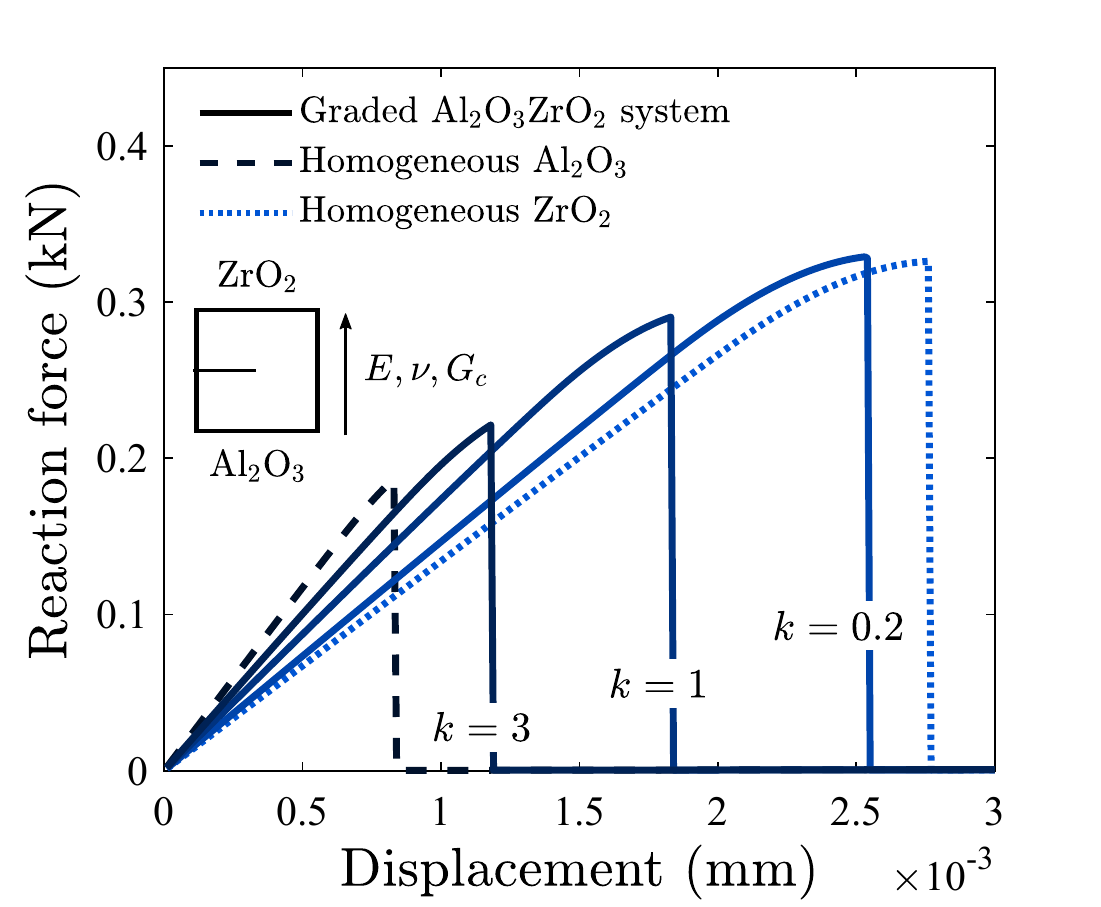}
                \caption{}
                \label{fig:Case1yaxis}
        \end{subfigure}
        \begin{subfigure}[b]{0.55\textwidth}
                \raggedleft
                \includegraphics[scale=0.7]{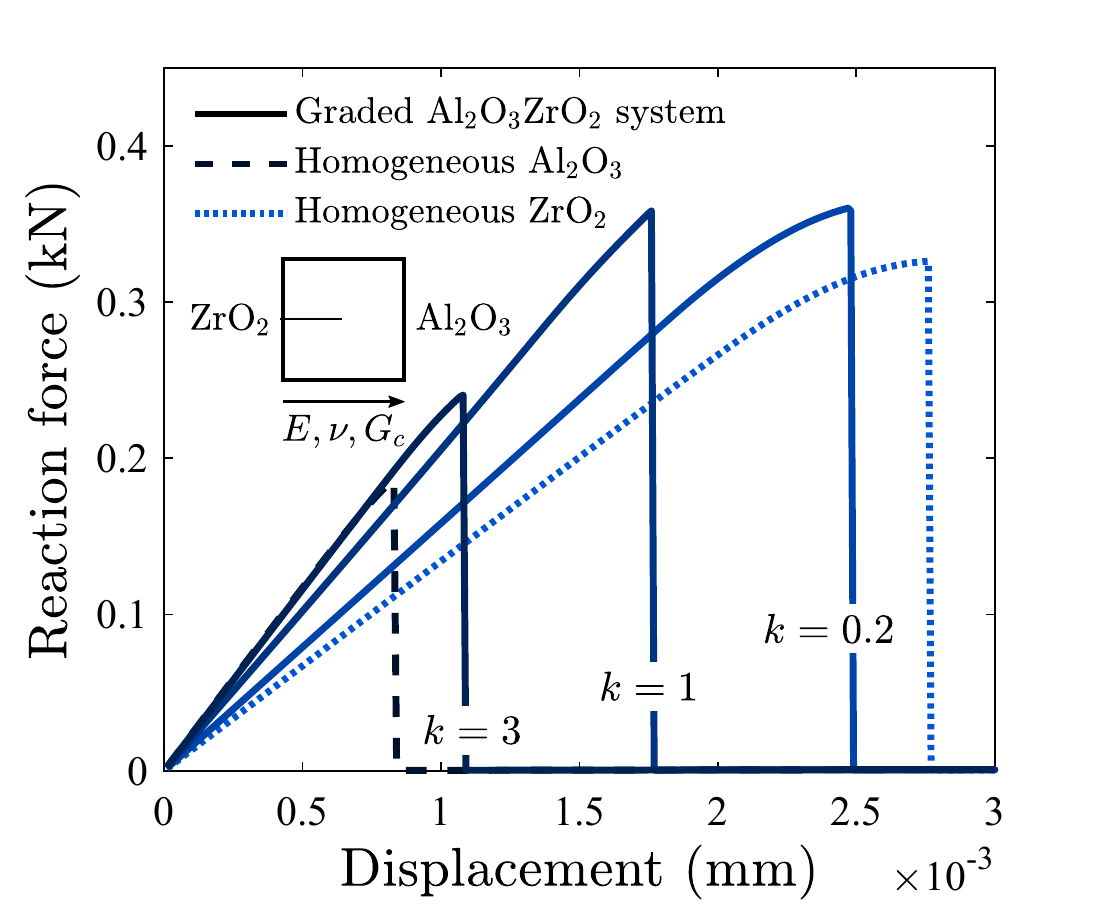}
                \caption{}
                \label{fig:Case1xaxisZrO2}
        \end{subfigure}}

\makebox[\linewidth][c]{%
        \begin{subfigure}[b]{0.55\textwidth}
                \centering
                \includegraphics[scale=0.7]{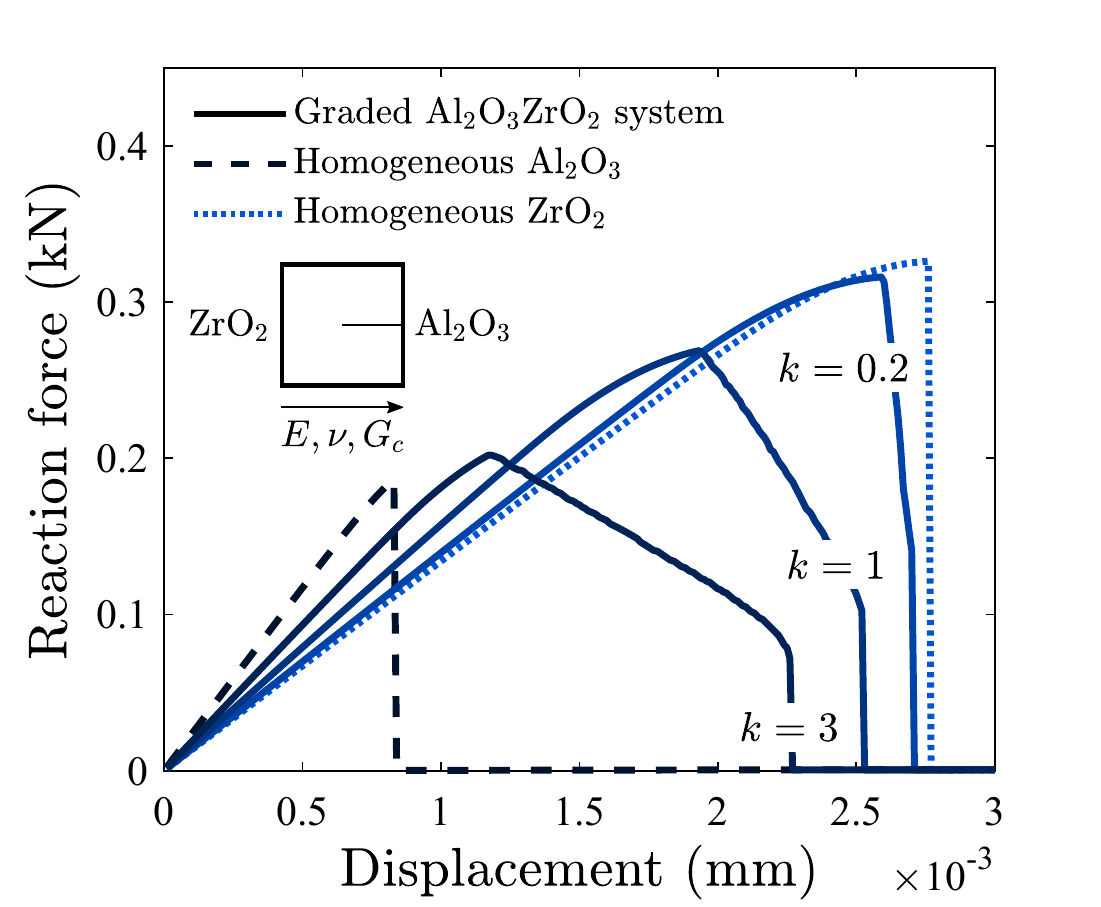}
                \caption{}
                \label{fig:Case1xaxisAl2O3}
        \end{subfigure}
        \begin{subfigure}[b]{0.55\textwidth}
                \raggedleft
                \includegraphics[scale=0.7]{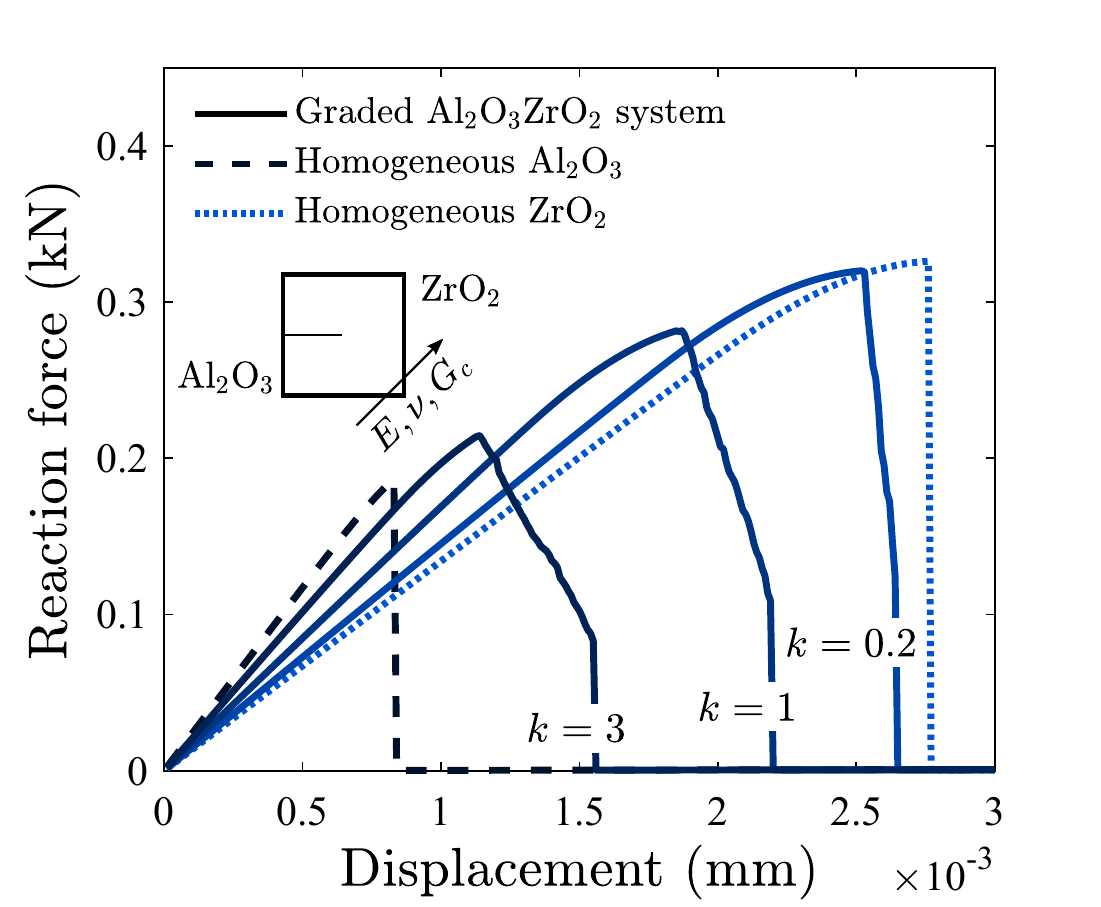}
                \caption{}
                \label{fig:Case145deg}
        \end{subfigure}
        }       
        \caption{Load versus displacement curves for material gradation along: (a) the $y$-axis, (b) the $x$-axis, with the initial crack on the ZrO$_2$ side, (c) the $x$-axis, with the initial crack on Al$_2$O$_3$ side, and (d) with an inclination of $45^\circ$ relative to the crack.}\label{fig:Case1}
\end{figure}

The influence of mode-mixity due to material gradients is investigated through four scenarios, with material properties varying: (1) along the $y$-axis, perpendicular to the crack; (2) along the $x$-axis, with the crack placed in the ZrO$_2$ edge; (3) along the $x$-axis, with the crack placed in the Al$_2$O$_3$ edge; and, (4) with an inclination of 45$^\circ$ degrees relative to the crack. The finite element results obtained are shown in Fig. \ref{fig:Case1}.

Consider first the results obtained for a material property gradient perpendicular to the crack, see Fig. \ref{fig:Case1yaxis}. As expected, see Table \ref{Tab:AlO3ZrO2}, alumina exhibits a stiffer response than zirconia but an earlier fracture. In the functionally graded cases, the elastic and fracture properties are constant along $x$ and the material gradient brings crack tip mode mixity. As inferred from Fig. \ref{fig:Case1matproperties}, the material along the crack path will have a larger value of $E$ with increasing $k$ but a smaller $G_c$. The force versus displacement response is consistent, with the curves showing an increased stiffness with increasing $k$ but a smaller fracture resistance. When the material gradient is parallel to the crack, Figs. \ref{fig:Case1xaxisZrO2} and \ref{fig:Case1xaxisAl2O3}, the response differs drastically depending on where the crack is placed. In all cases the FGM results render a stiffness slope and a maximum force that lie between the ones obtained from the homogeneous samples. In addition, the same qualitative response is observed when increasing $k$; as in Fig. \ref{fig:Case1yaxis}, the stiffness increases but failure occurs at a lower load level. However, the response after crack initiation is very different. When the crack propagates from the zirconia-rich side to the alumina-rich side, the value of $G_c$ decreases along the crack path and a sharp drop in the load-displacement curve is observed, see Fig. \ref{fig:Case1xaxisZrO2}. On the other hand, when the initial crack lies in the alumina-rich side, Fig. \ref{fig:Case1xaxisAl2O3}, the FGM specimen exhibits an increased resistance to crack propagation. The crack progressively propagates into material with an increasing $G_c$ and the remote displacement required to cause complete breakage can duplicate the displacement attained at the peak load. This enhanced resistance to crack growth persists when the material gradient is not aligned with the crack flank, as observed in the results of Fig. \ref{fig:Case145deg}. It is therefore shown that FGMs can be tailored to gain toughness resistance and prevent unstable fracture, even in the case of brittle ceramic-based FGMs.
  
\subsection{Mixed mode fracture of a graded photodegradable copolymer}
\label{Sec:CaseStudy2}

\begin{figure}[H]
        \centering
        \begin{subfigure}[h]{0.49\textwidth}
                \centering
                \includegraphics[scale=0.55]{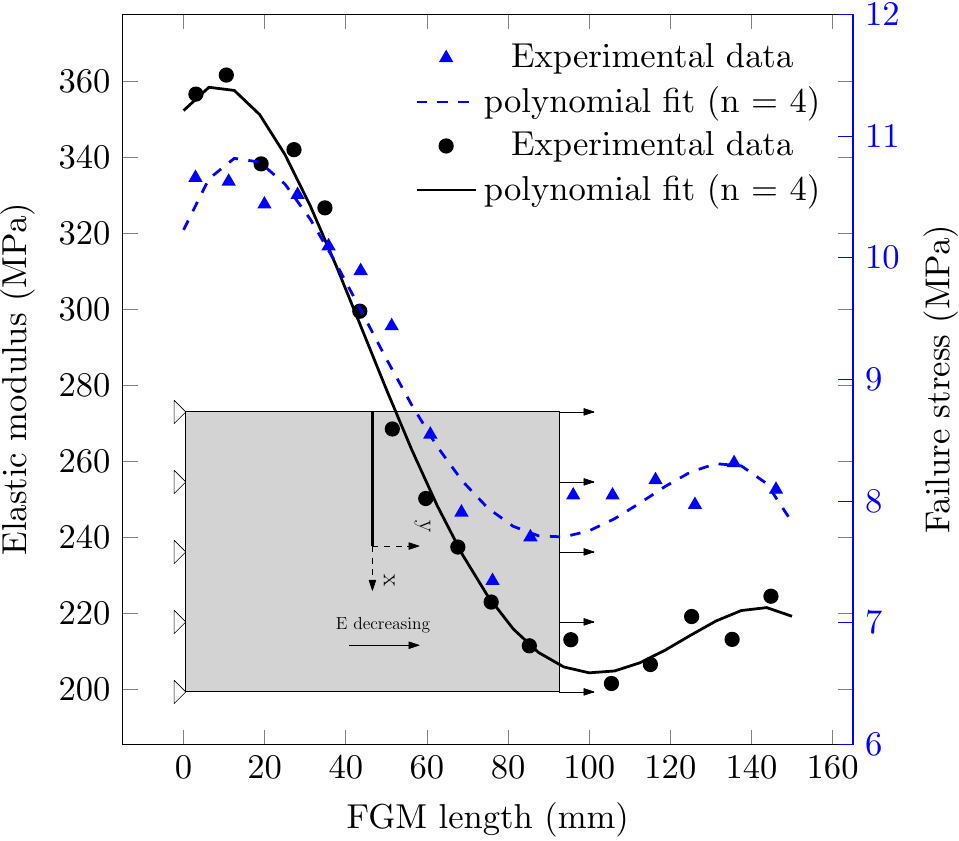}
                \caption{}
                \label{fig:m_case1}
        \end{subfigure}
        \begin{subfigure}[h]{0.49\textwidth}
                \centering
                \includegraphics[scale=0.55]{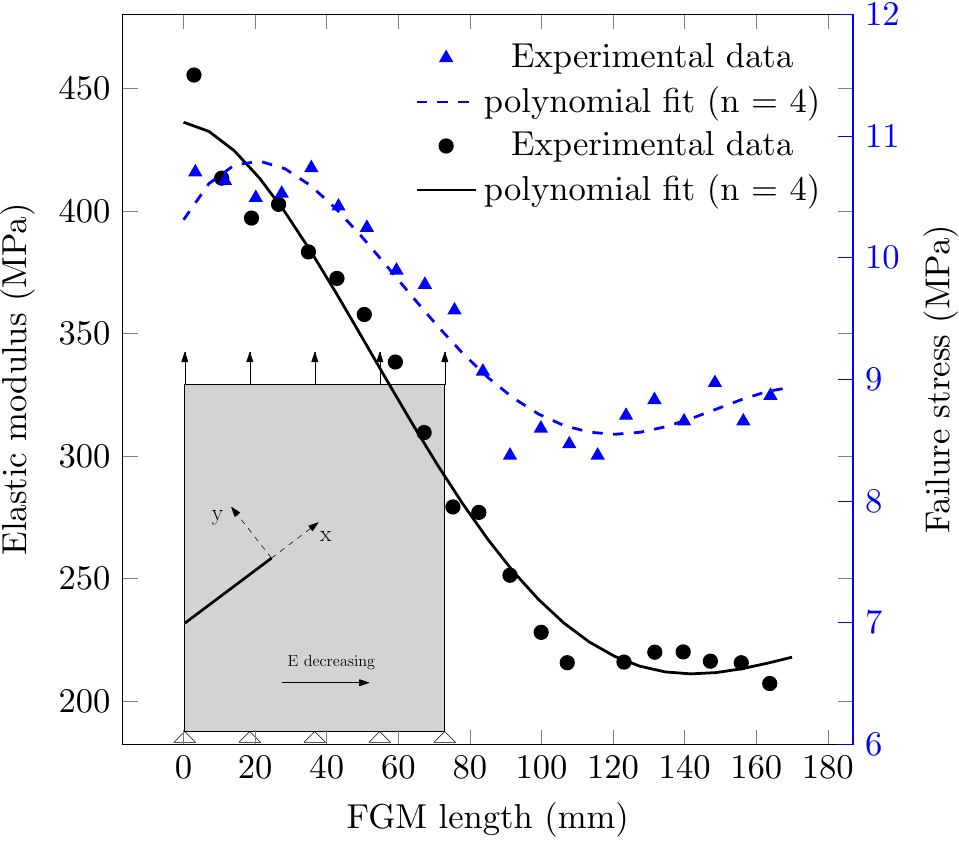}
                \caption{}
                \label{fig:m_case2}
        \end{subfigure}\\
        \vspace{10pt}
        \begin{subfigure}[h]{0.5\textwidth}
                \centering
                \includegraphics[scale=0.55]{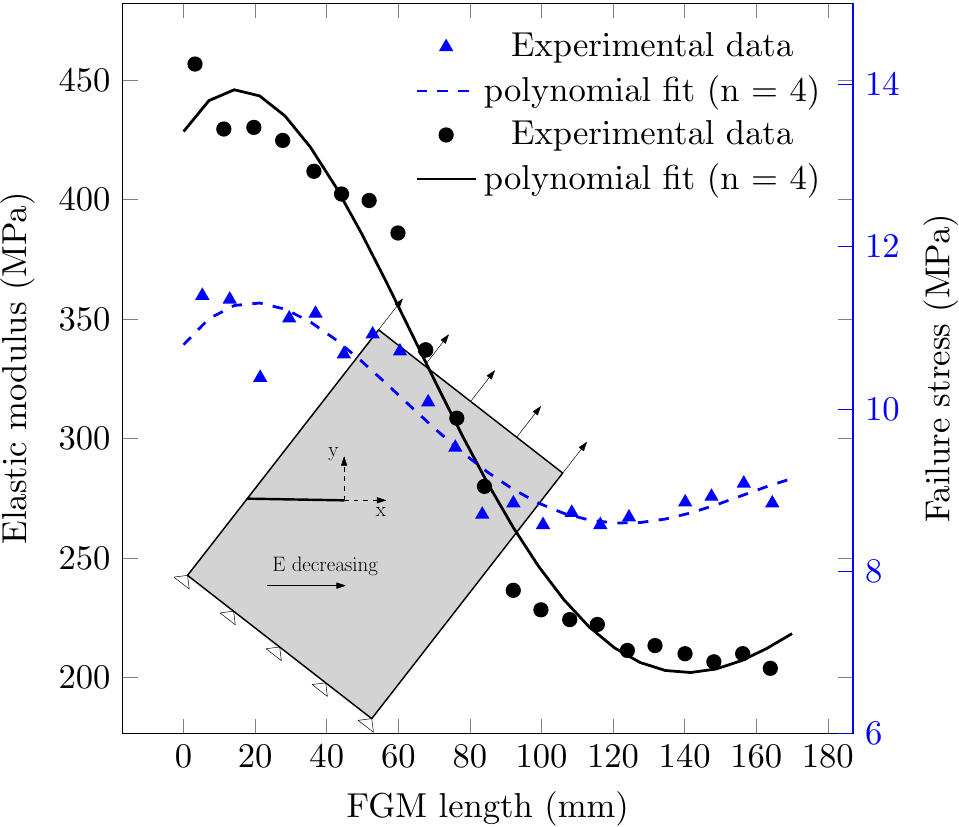}
                \caption{}
                \label{fig:m_case3}
        \end{subfigure}
        \caption{Geometry and boundary conditions: (a) FGM I, crack perpendicular to material gradient, (b) FGM II, crack at an angle relative to remote loading, and (c) FGM III, crack at an angle relative to remote loading and material gradient at an angle relative to the mode I crack path.}\label{fig:case3_materialproperties}
\end{figure}

\begin{figure}[H]
        \centering
        \begin{subfigure}[h]{0.4\textwidth}
                \centering
                \includegraphics[scale=0.375]{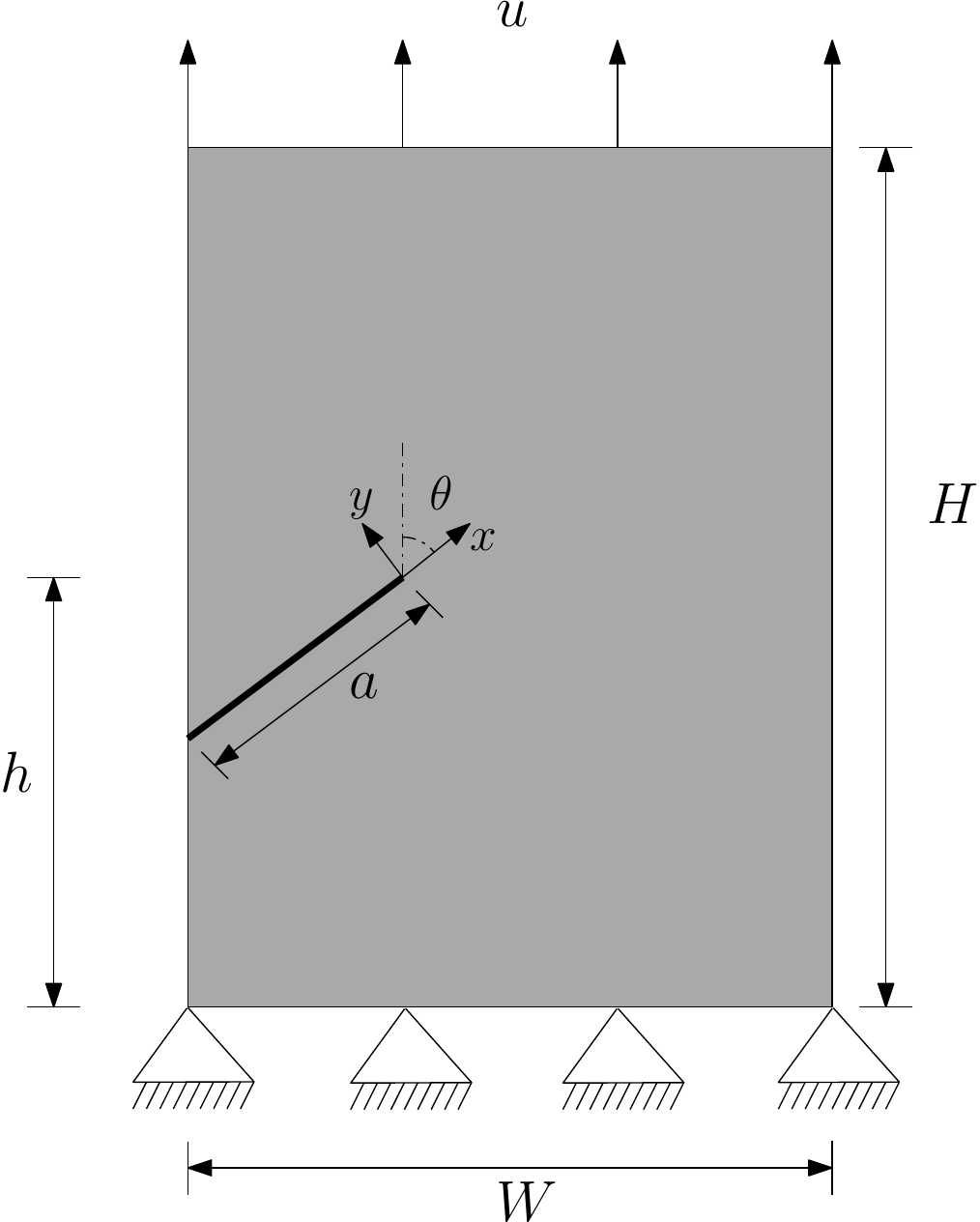}
                \caption{}
                \label{fig:case2_domain_homogeneous}
        \end{subfigure}
        \begin{subfigure}[h]{0.4\textwidth}
                \centering
                \includegraphics[scale=0.15]{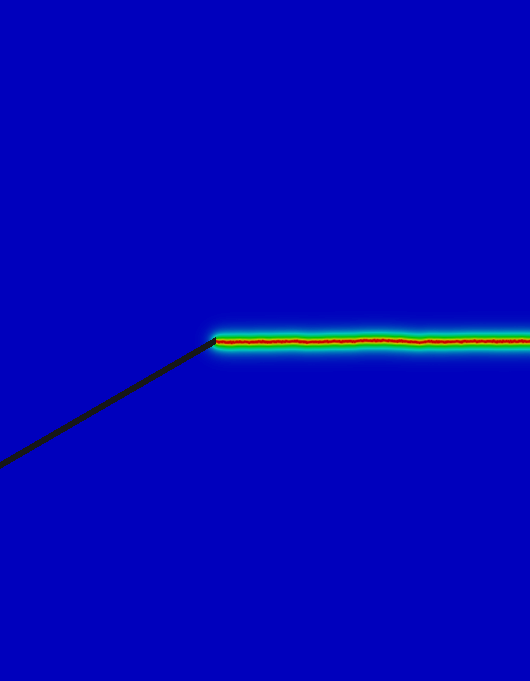}
                \caption{}
                \label{fig:case2_crack_homogeneous}
        \end{subfigure}
        \caption{Edge cracked plate: (a) geometry and boundary conditions, and (b) crack path predicted for the homogeneous case.}\label{fig:case2}
\end{figure}

We assess the capabilities of the modelling framework in capturing mixed-mode crack propagation in FGMs. While it is widely considered that the phase field fracture method holds great promise in dealing with crack propagation under mixed-mode conditions, even in homogeneous material comparisons with experiments are scarce \cite{Pham2017}. Here, model predictions are benchmarked against the experimental measurements by Abanto-Bueno and Lambros \cite{Abanto-Bueno2006a} on polyethylene cocarbon monoxide (ECO) exposed to selective UV irradiation. ECO undergoes accelerated mechanical degradation when exposed to UV light so that a sample with continuous in-plane property gradation can be obtained by gradually irradiating a sheet of the material. The material gradation profile is characterized, in terms of Young's modulus $E$ and failure stress $\sigma_f$, by cutting small strips that are then subjected to uniaxial tension testing. As shown in Fig. \ref{fig:case3_materialproperties}, we choose to fit the experimental data points with a fourth order polynomial to define the variation of $E$ and $\sigma_f$. The variation of the critical energy release rate is calculated from the $\sigma_f$ measurements by making use of the stress intensity factor solution $K_I= Y \sigma \sqrt{\pi a}$, where $Y$ is the geometry factor. As shown in Fig. \ref{fig:case2}, the geometry under consideration is an edge cracked plate subjected to uniaxial tension. Three different configurations are considered so as to explore the three characteristic
geometries of mixed-mode fracture in FGMs; crack tip mixity can be attained either by placing the crack at an angle to the direction of material property variation, or by asymmetric external loading (as in the homogeneous case), or by a combination of both. The effect of each of these cases
is here investigated using three specimens respectively labeled FGM I, II, and III. The geometry and dimensions of the specimens are shown in Table \ref{table:case3_geometry} and Figs. \ref{fig:case3_materialproperties} and \ref{fig:case2}, along with the data for the reference homogeneous case. In all cases Poisson's ratio is taken to be constant throughout the specimen and equal to $\nu=0.45$, in agreement with the experimental measurements, see \cite{Abanto-Bueno2006a} and references therein. In the homogeneous case, the material properties read $E=280$ MPa and $G_c=27$ J/m$^2$. The phase field length scale is taken to be equal to 0.7 mm. After a mesh sensitivity study, 191735 linear triangular elements are employed to discretize the geometry.

\begin{table}[H]
\centering
\caption{Dimensions, in mm, of the ECO specimens.}
\label{table:case3_geometry}
\begin{tabular}{lccccc}
\hline
       & H & W & h & a & $\theta~ (rad)$ \\ \hline
Homogeneous   & 90     & 70     & 45  & 33    & $\pi/3$        \\
FGM I   & 75     & 70     & 37.5  & 30    & $\pi/2$        \\
FGM II  & 90     & 70     & 32    & 26    & $\pi/3$        \\
FGM III & 90     & 70     & 32    & 25    & $\pi/3$        \\ \hline
\end{tabular}
\end{table}

\begin{figure}[H]
\makebox[\linewidth][c]{%
        \begin{subfigure}[b]{0.5\textwidth}
                \centering
                \includegraphics[scale=0.5]{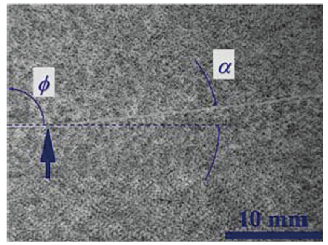}
                \caption{}
                \label{fig:ExptCrackI}
        \end{subfigure}
        \begin{subfigure}[b]{0.5\textwidth}
                \centering
                \includegraphics[scale=0.58]{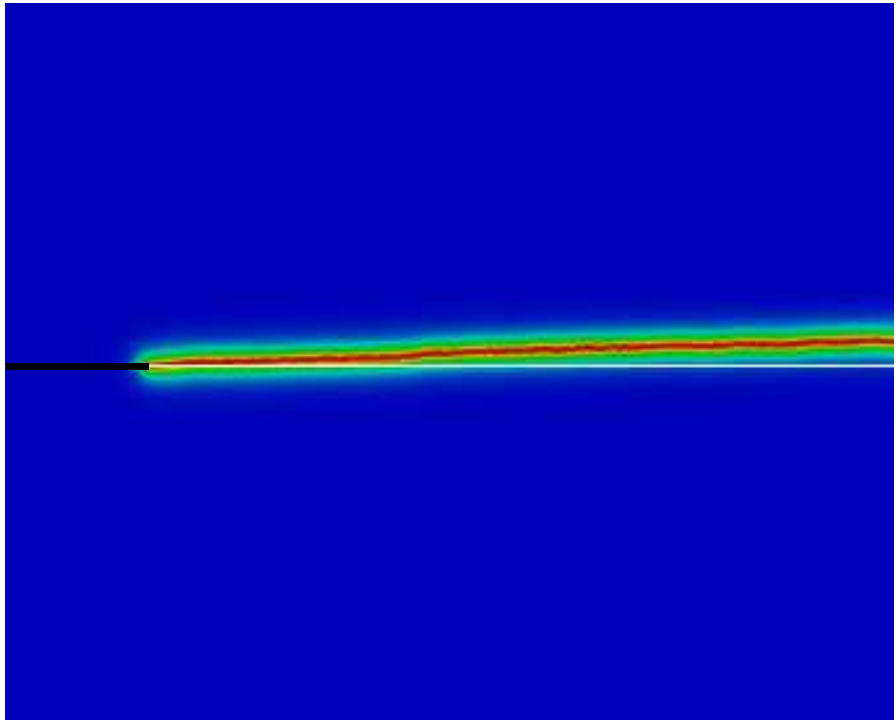}
                \caption{}
                \label{fig:NumCrackI}
        \end{subfigure}}

\makebox[\linewidth][c]{%
        \begin{subfigure}[b]{0.5\textwidth}
                \centering
                \includegraphics[scale=0.5]{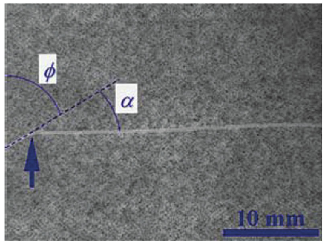}
                \caption{}
                \label{fig:ExptCrackII}
        \end{subfigure}
        \begin{subfigure}[b]{0.5\textwidth}
                \centering
                \includegraphics[scale=0.42]{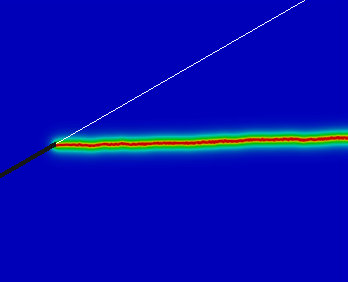}
                \caption{}
                \label{fig:NumCrackII}
        \end{subfigure}}   
        
\makebox[\linewidth][c]{%
        \begin{subfigure}[b]{0.5\textwidth}
                \centering
                \includegraphics[scale=0.5]{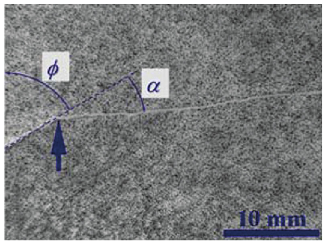}
                \caption{}
                \label{fig:ExptCrackIII}
        \end{subfigure}
        \begin{subfigure}[b]{0.5\textwidth}
                \centering
                \includegraphics[scale=0.65]{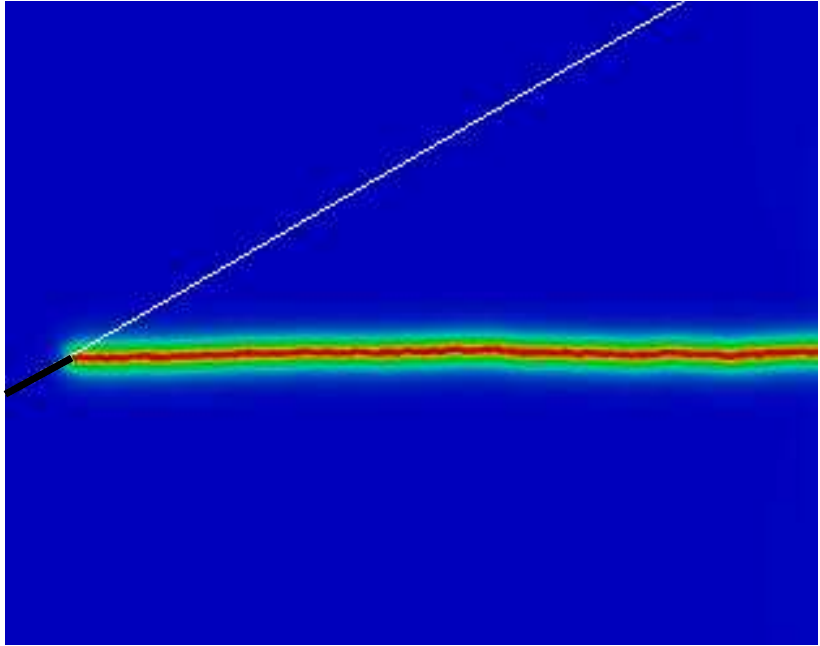}
                \caption{}
                \label{fig:NumCrackIII}
        \end{subfigure}}          
        \caption{Numerical predictions of crack propagation paths and comparison with experiments \cite{Abanto-Bueno2006a}: (a) and (b), FGM I - crack perpendicular to material gradient; (c) and (d), crack at an angle relative to the remote load; and (e) and (f), mode-mixity due to crack inclination and material gradient.}\label{fig:case2_crackpropagation}
\end{figure}

The finite element results obtained are shown in Table \ref{table:case2_crack_kinking} and Fig. \ref{fig:case2_crackpropagation}, along with the experimental measurements by Abanto-Bueno and Lambros \cite{Abanto-Bueno2006a}. Initial crack kinking angle predictions are in very close agreement with the experimental measurements (Table \ref{table:case2_crack_kinking}). The phase field method predictions exhibit a better correlation with the experiments than other criteria, such as the so-called maximum tangential stress (MTS) or the generalized maximum tangential stress (GMTS), see \cite{Oral2008a}. 

\begin{table}[H]
\centering
\caption{Experimental and numerical predictions of crack initiation angles.}
\label{table:case2_crack_kinking}
\begin{tabular}{ccc}
\hline
\multicolumn{3}{c}{Crack initiation angles, $\alpha$(deg)} \\ \hline
\multicolumn{2}{r}{Experiment} & Numerical model \\ \hline
Homogeneous          & -28 $\pm$ 1.5          & -30.4          \\
FGM I           & -0 $\pm$ 1.5          & 1.4          \\ 
FGM II          & -28 $\pm$ 1.5         & -29.9         \\
FGM III         & -26 $\pm$ 1.5         & -27.6            \\ \hline
\end{tabular}
\end{table}

The resulting crack propagation paths are shown in Fig. \ref{fig:case2_crackpropagation}. Again, a good agreement with the experiments is observed. In the case where the crack is perpendicular to the material gradient, FGM I - Figs. \ref{fig:ExptCrackI} and \ref{fig:NumCrackI}, the crack deviates from the natural trajectory of the first mode of fracture and propagates towards the more compliant part of the specimen, where the fracture resistance is also diminishing. For the case FGMII, numerical predictions capture the initial crack kink due to crack tip mode mixity, and the crack then propagates following the distinctive path inherent to the first mode of fracture (see Figs. \ref{fig:ExptCrackII} and \ref{fig:NumCrackII}). In FGM III the crack tip axisymmetric stress field induces crack deflection, which is further exacerbated by the $G_c$ spatial gradient. The numerical result qualitatively captures this trend, see Figs. \ref{fig:ExptCrackIII} and \ref{fig:NumCrackIII}.

\subsection{Failure of compositionally graded glass-filled epoxy specimens}
\label{Sec:CaseStudy3}

\begin{figure}[H]
        \centering
        \begin{subfigure}[h]{1.\textwidth}
                \centering
                \includegraphics[scale=0.8]{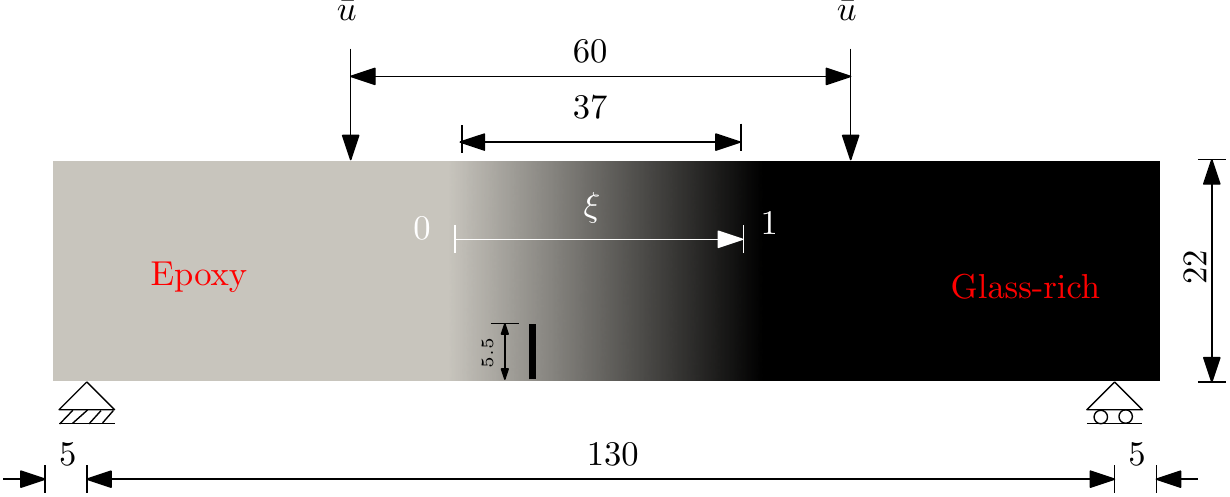}
                \caption{}
                \label{fig:4PBending}
        \end{subfigure}\\
        \begin{subfigure}[h]{1.\textwidth}
                \centering
                \includegraphics[scale=0.8]{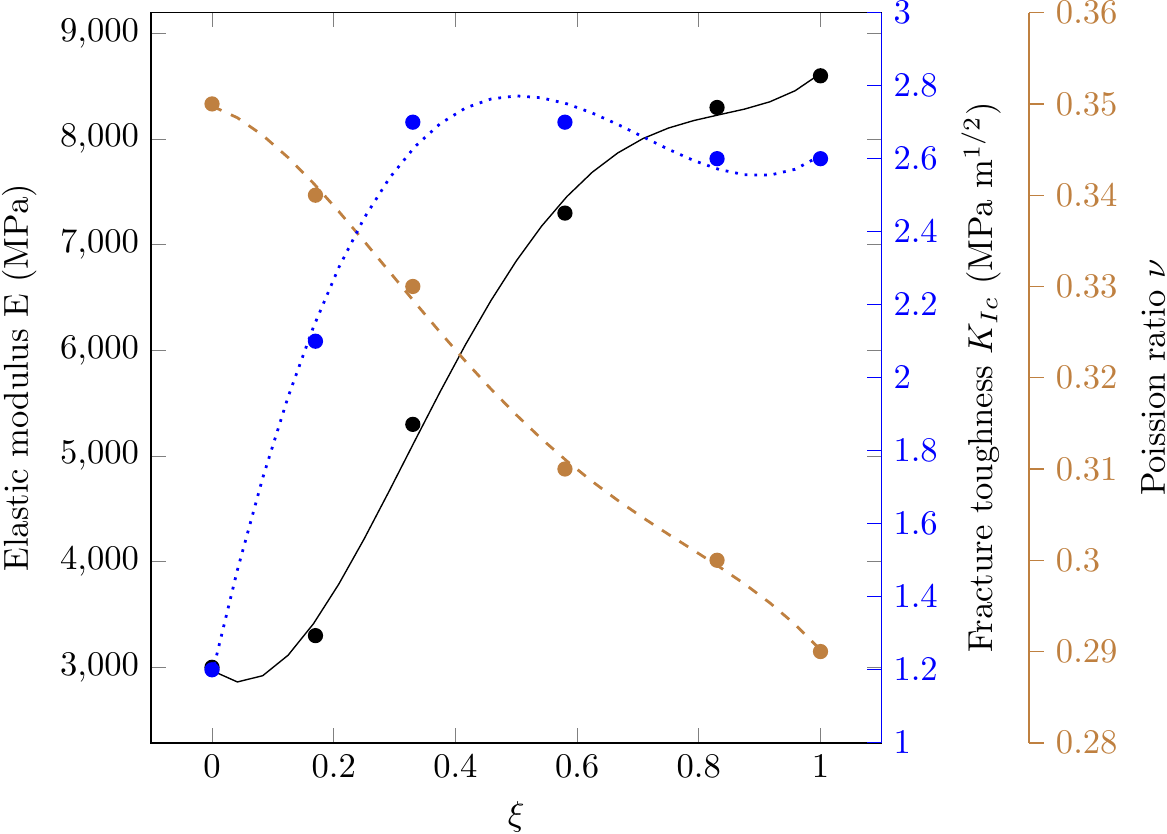}
                \caption{}
                \label{fig:BendingMatproperties}
        \end{subfigure}
        \caption{ Four point bending of epoxy-glass FGMs, (a) geometry and boundary conditions (dimensions in mm), and (b) material properties variation.}\label{fig:domain_fourPoint}
\end{figure}

We further assess the capabilities of the model in predicting experimental results by capturing crack growth in graded glass-filled epoxy specimens. We aim at addressing a wide range of FGMs by benchmarking with samples resulting from different manufacturing and characterization techniques. Phase field predictions are compared with the four point bending experiments by Rousseau and Tippur \cite{Rousseau2000}. As shown in Fig. \ref{fig:4PBending}, material properties change gradually from the epoxy side to the glass-rich side. The length of the graded domain is normalized by $\xi$, which goes from 0, 100\% epoxy, to 1, 100\% glass. Three tests are conducted, with the crack being placed at different locations: $\xi= 0.17, 0.58, 1$. The elastic property variation is measured by using ultrasonic pulse-echo measurements; the material property gradient is shown in Fig. \ref{fig:BendingMatproperties}. The length scale equals $\ell=0.524$ mm and the variation of $G_c$ is obtained from the fracture toughness measurements. The finite element model reproduces the experimental setup, and a total of 90917 triangular elements are employed.\\

\begin{figure}[H]
        \centering
        \begin{subfigure}[h]{0.32\textwidth}
                \centering
                \includegraphics[scale=0.36]{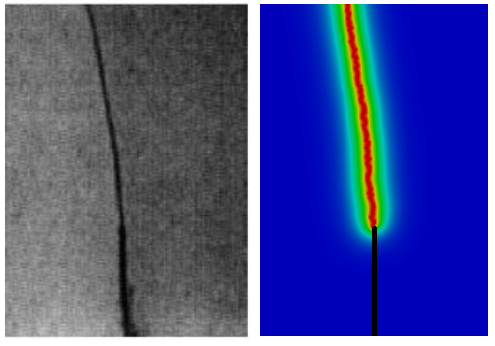}
                \caption{}
                \label{fig:crackpath1}
        \end{subfigure}
        \begin{subfigure}[h]{0.32\textwidth}
                \centering
                \includegraphics[scale=0.56]{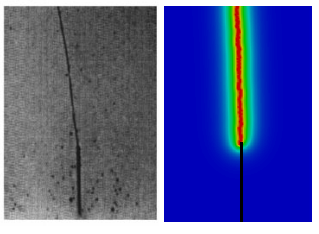}
                \caption{}
                \label{fig:crackpath2}
        \end{subfigure}
        \begin{subfigure}[h]{0.32\textwidth}
                \centering
                \includegraphics[scale=0.56]{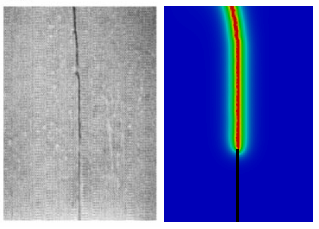}
                \caption{}
                \label{fig:crackpath3}
        \end{subfigure}
        \caption{Epoxy-glass FGMs. Experiments \cite{Rousseau2000} and numerical predictions of crack trajectories for initial crack locations (a) $\xi=0.17$, (b) $\xi=0.58$, and (c) $\xi=1.0$.}\label{fig:crack_kinking_fourPoint}
\end{figure}

The crack propagation paths predicted are shown in Fig. \ref{fig:crack_kinking_fourPoint} along with their experimental counterparts for each initial crack location. A very good agreement with the experiments is observed. As the crack location moves towards the epoxy-rich side, crack deflection is enhanced. The crack propagates into the more compliant part, as this results in a greater release of elastic energy. Furthermore, the fracture resistance reduces with diminishing $\xi$, contributing further to the deviation of the crack path trajectory from the conventional mode I path.

\begin{figure}[H]
    \centering
    \includegraphics[scale = 1.0]{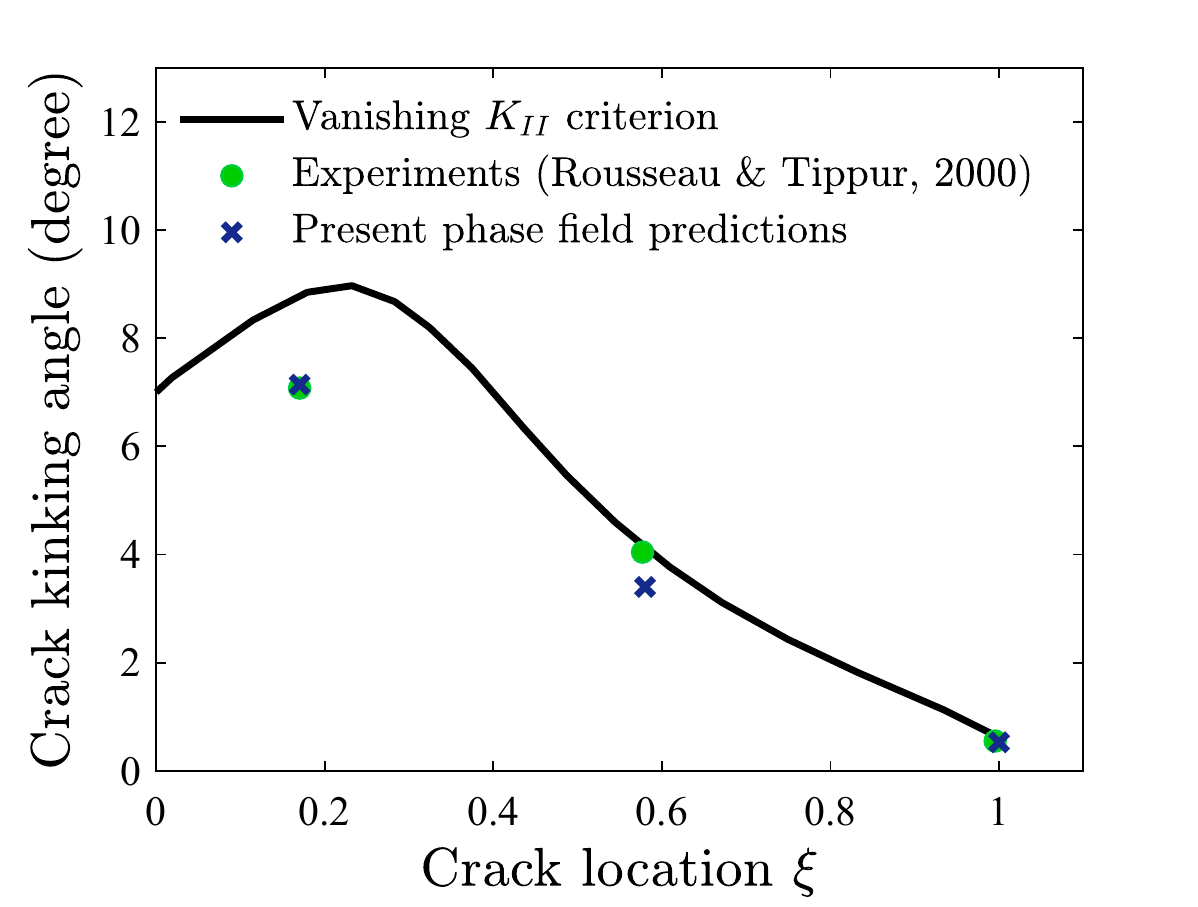}
    \caption{Epoxy-glass FGMs. Crack kinking angles as a function of the initial crack location along the graded region, $\xi$.}
    \label{fig:crack_kinking_fourPoint_angle}
\end{figure}

The initial crack deflections, or crack kinking angles, are shown in Fig. \ref{fig:crack_kinking_fourPoint_angle}. We compare our phase field predictions with the experimental measurements by Rousseau and Tippur \cite{Rousseau2000} and with the semi-analytical solution given by the vanishing $K_{II}$ criterion. As shown in Fig. \ref{fig:crack_kinking_fourPoint_angle}, a very good agreement with the experimental measurements is attained, improving the predictions of the vanishing $K_{II}$ criterion. The results emphasize the relevance of capturing the spatial variation of $G_c$ to attain accurate predictions; $K_{II}$ vanishes at the direction of maximum energy release rate.

\subsection{Complex crack patterns in functionally graded solids}
\label{Sec:CaseStudy4}

\begin{figure}[H]
    \centering
    \includegraphics[scale = 0.45]{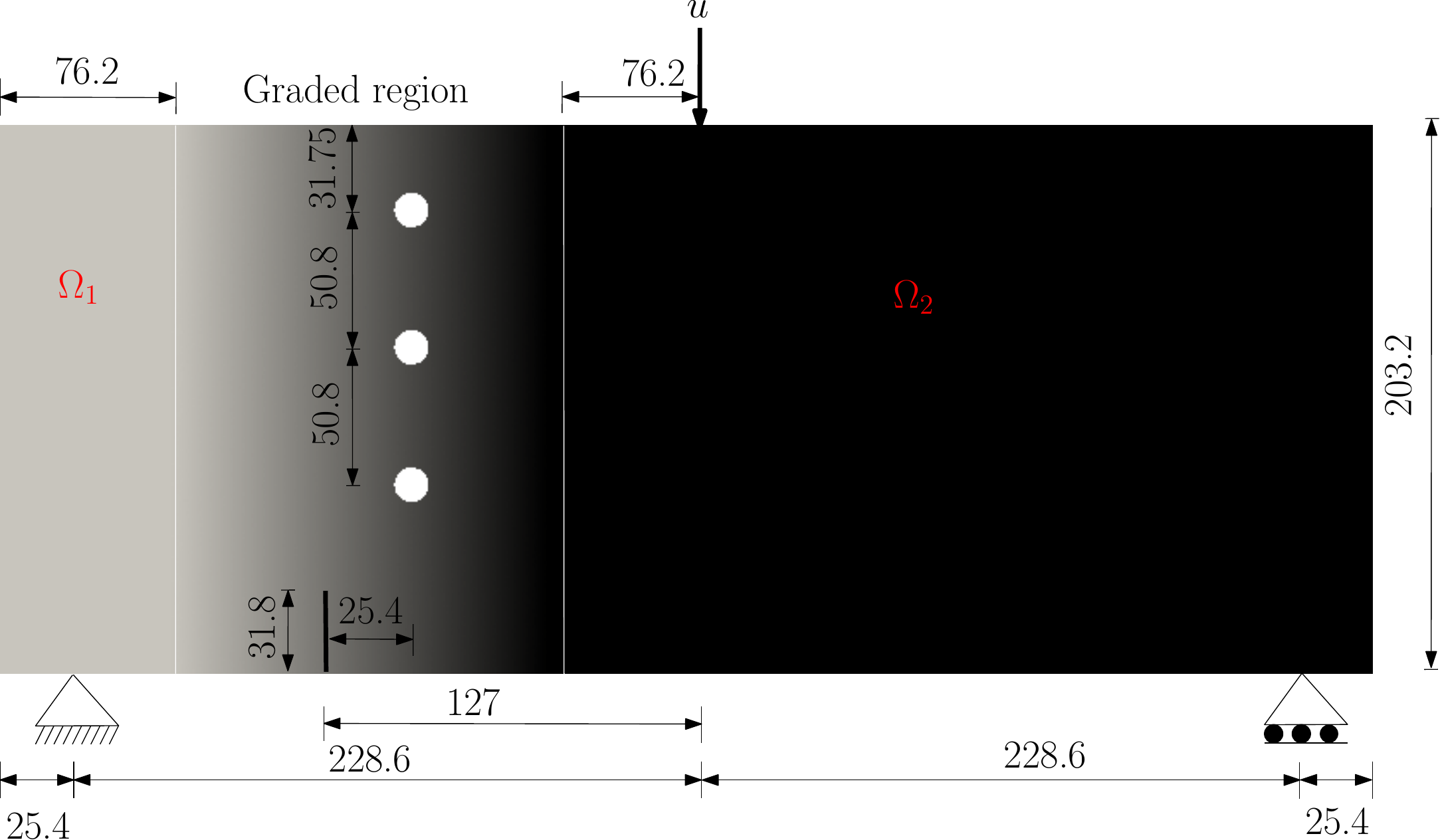}
    \caption{Geometry and loading configuration, all dimensions are given in mm. The radius of the holes equals 6.35 mm.}
    \label{fig:threeHoles_threePoint_BC}
\end{figure}

The capabilities of the present phase field formulation for FGMs in predicting complex crack propagation paths are assessed and compared to results from other numerical techniques. We choose to simulate the paradigmatic problem of a cracked beam with three holes, see Fig. \ref{fig:threeHoles_threePoint_BC}. The holes are misaligned relative to the initial crack, deviating the crack trajectory from the mode I path. In addition, the material gradient further contributes to the crack trajectory. Material properties vary linearly over the graded region from a domain $\Omega_1$, with material properties $E=1$ MPa, $\nu=0.3$ and $K_{Ic}=1$ MPa$\sqrt{m}$, to a domain $\Omega_2$, with $E=3$ MPa, $\nu=0.3$ and $K_{Ic}=1.5$ MPam$\sqrt{m}$. The non-homogeneous beam is subjected to three point bending, as outlined in Fig. \ref{fig:threeHoles_threePoint_BC}. The entire specimen is discretized by means of 236253 triangular elements so as to resolve a phase field length scale of $\ell$ = 1 mm along the potential crack path.

\begin{figure}[H]
        \centering
        \begin{subfigure}[h]{0.23\textwidth}
                \centering
                \includegraphics[scale=0.23]{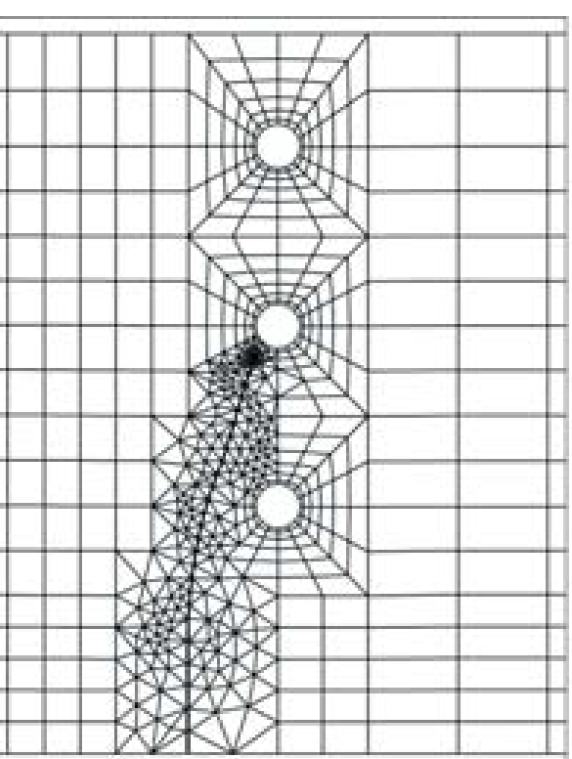}
                \caption{}
                \label{fig:KimPaulino}
        \end{subfigure}
        \begin{subfigure}[h]{0.3\textwidth}
                \centering
                \includegraphics[scale=0.45]{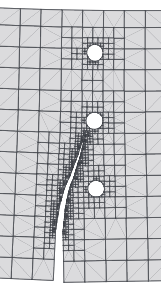}
                \caption{}
                \label{fig:Chenetal}
        \end{subfigure} \hspace{-12mm}
        \begin{subfigure}[h]{0.32\textwidth}
                \centering
                \includegraphics[scale=0.23]{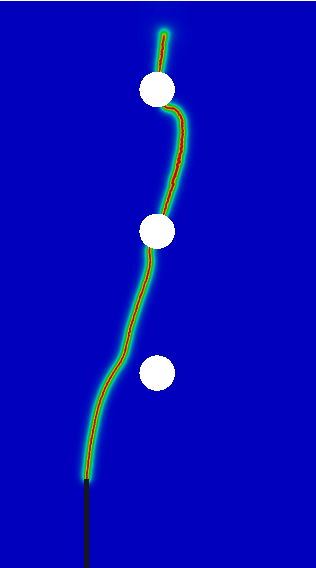}
                \caption{}
                \label{fig:Ours}
        \end{subfigure}
        \caption{Complex crack path predictions in functionally graded solids with (a) finite element re-meshing techniques \cite{Kim2004}, (b) scaled boundary finite element methods \cite{Chen2018}, and (c) present phase field fracture predictions.}\label{fig:Case4comparison}
\end{figure}

The results obtained, in terms of the crack trajectory, are shown in Fig. \ref{fig:Case4comparison}. The phase field predictions are compared with the results obtained by (a) Kim and Paulino \cite{Kim2004}, using the maximum energy release rate criterion and a finite element re-meshing algorithm, and (b) Chen et al. \cite{Chen2018}, with the maximum circumferential stress criterion and the scaled boundary finite element method. A similar crack path is predicted in the three cases; the crack deflects towards the hole region, eventually coalescing with the intermediate hole. The phase field method has the capacity to predict crack nucleation, and consequently the full trajectory until complete separation can be predicted; the crack re-emerges at the top of the intermediate hole, and coalesces with the third hole before reappearing again. Another advantage of the present formulation is the capability of predicting the crack propagation path with the initial mesh. However, if the crack trajectory is uncertain, combining the phase field method with a re-meshing algorithm may prove more computationally efficient.

\subsection{3D crack propagation in functionally graded plates}
\label{Sec:CaseStudy5}

\begin{figure}[H]
    \centering
    \includegraphics[scale=0.75]{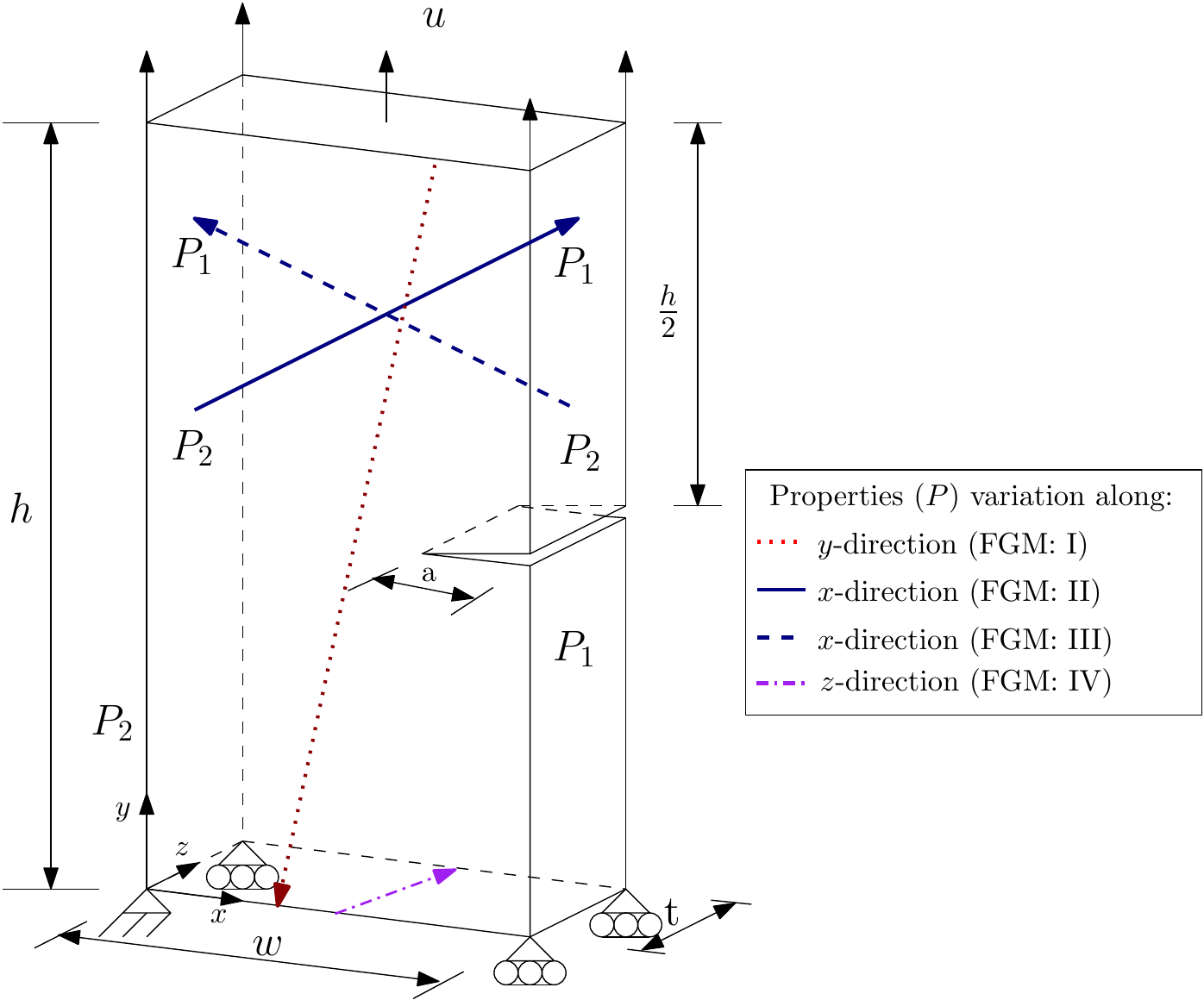}
    \caption{Geometry, material and loading configuration in a 3D functionally graded specimen. Four cases are considered based on the variatiation of the elastic and fracture properties.}
    \label{fig:3D_Domain}
\end{figure}

The present modelling framework can also be used to address large-scale engineering problems, as well as to gain insight into the design of three dimensional FGMs with enhanced fracture properties. We demonstrate such capabilities by modelling a plate subjected to uniaxial tension with an edge crack through the thickness, see Fig. \ref{fig:3D_Domain}. Four combinations of crack versus material gradient orientation are considered. We name FGM I the case where the elastic and fracture properties vary gradually along the vertical $y$-axis. In the case of a material gradient along the horizontal axis, we consider two cases (FGM II and FGM III), depending on the location of the crack relative to the compounds. In addition, we consider the case where the material gradient varies along the thickness direction ($z$-axis in Fig. \ref{fig:3D_Domain}); FGM IV. For simplicity, we assume a linear variation of material properties between two cases: material A, with Young's modulus $E=300$ GPa, critical energy release rate $G_c=0.85$ N/mm, and Poisson's ratio $\nu=0.25$; and material B, with $E=120$ GPa, $G_c=1.15$ N/mm, $\nu=0.35$. A constant length scale is assumed, $\ell=0.031$ mm. In FGM II the crack is located on the B-rich edge. A homogeneous case will also be considered, with properties $E=210$ GPa, $\nu=0.3$ and $G_c=1$ N/mm. The specimen, outlined in Fig. \ref{fig:3D_Domain}, has width $w=1$ mm, height $h=2$ mm, thickness $t=0.2$ mm and initial crack length $a=0.1$ mm. The finite element model involves 359276 linear tetrahedron elements. Each simulation takes approximately 25 hours running on 2 cores.

\begin{figure}[H]
        \centering
        \begin{subfigure}[h]{0.35\textwidth}
                \raggedright
\begin{tikzpicture}[]
\draw (15, 0) node[inner sep=0] {    \includegraphics[scale = 0.25]{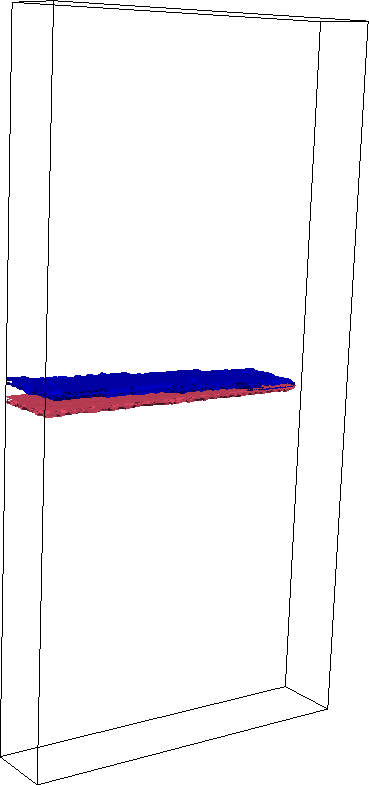}\label{fig:crack3d}};
 \draw [black](13.8, 0.6) node {\scriptsize {FGM I}}; 
 \draw [black](14.7, -0.65) node {\scriptsize {Homogeneous}};
 \draw[black,thick,->] (14,0.45) -- (14,0.14);
 \draw[black,thick,->] (14,-0.5) -- (14,-0.19);
 \draw[black,thick,-] (13.2,0.1) -- (13.2,-0.13);
 \draw[black,thick,-] (13.,0.1)--(13.35,0.1);
 \draw[black,thick,-] (13.,-0.1)--(13.35,-0.1);
\draw[black, thick, <-] (13.2,-0.1)--(13.2,-0.3);
\draw[black, thick, <-] (13.2,0.1)--(13.2,0.3);
  \coordinate (P) at (13.2,0.9);
\node[rotate=90] (N) at (P) {\scriptsize {0.05 mm}};  
\end{tikzpicture}
\caption{}
                \label{fig:3dcrackcontours}
        \end{subfigure}
        \begin{subfigure}[h]{0.64\textwidth}
                \centering
                \includegraphics[scale = 0.75]{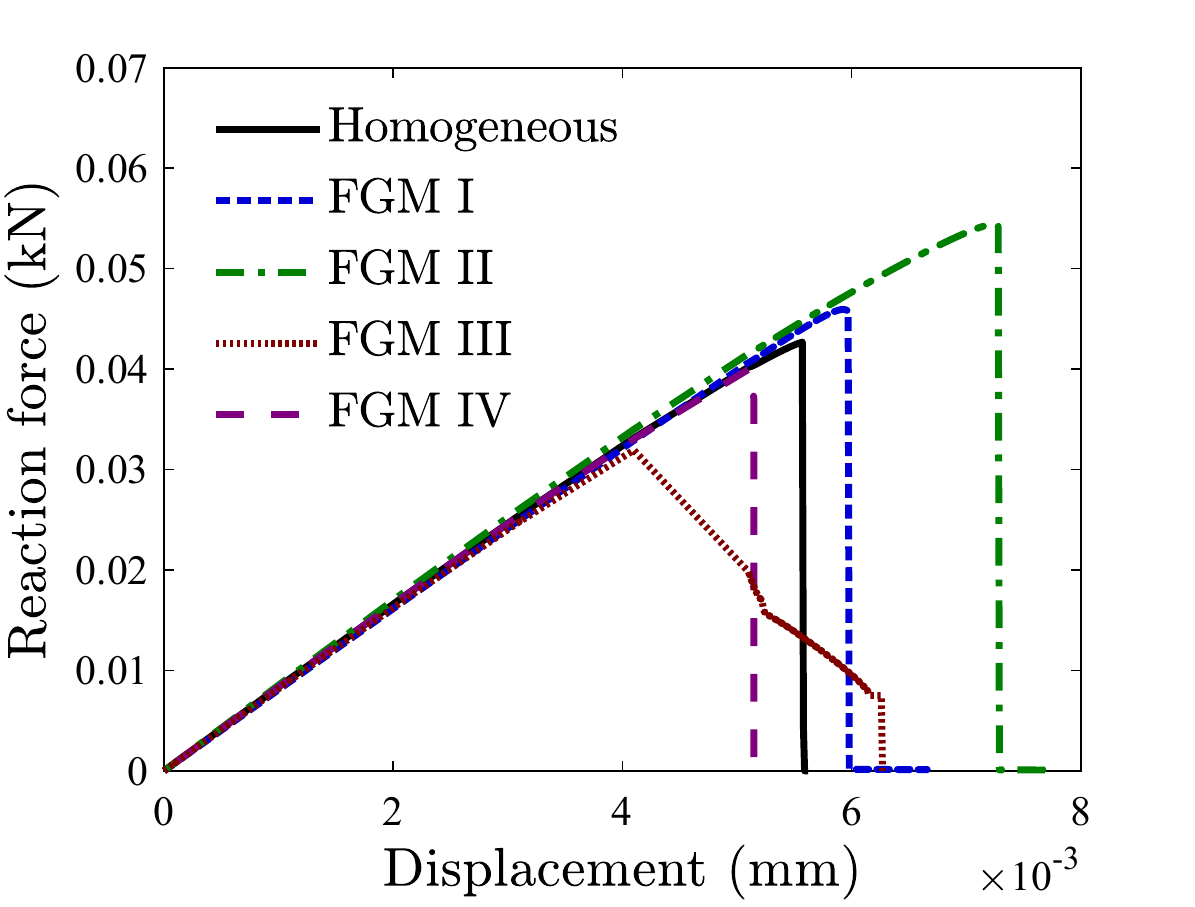}
                \caption{}
                \label{fig:3dforcevsDisp}
        \end{subfigure}
        \caption{Three dimensional FGM plate: (a) representative crack trajectories, and (b) load versus displacement curves.}\label{fig:threeDEdge}
\end{figure}
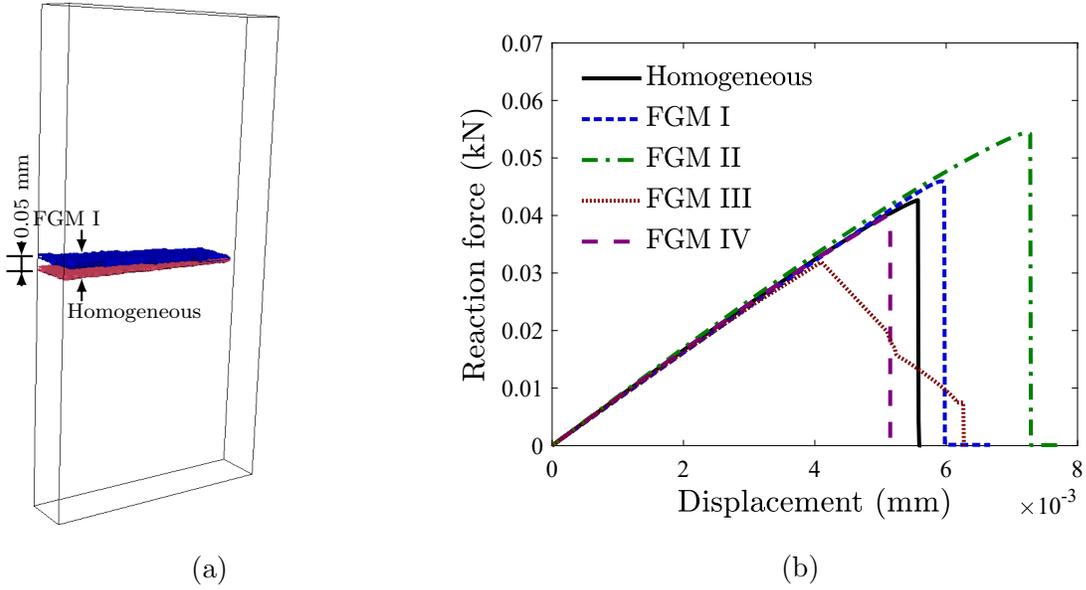

Results are shown in Fig. \ref{fig:threeDEdge} in terms of the force versus displacement response and representative crack paths. As shown in Fig. \ref{fig:3dcrackcontours}, the crack propagates along the extended crack plane in the homogeneous specimen but deviates towards the more compliant side in the graded one, where material properties vary vertically (FGM I). The load-displacement curves exhibits a similar response in both cases, but the graded sample delays fracture due to the local value of $G_c$ and the mode-mixity inherent to FGMs, see Fig. \ref{fig:3dforcevsDisp}. 

\begin{figure}[H]
        \centering
        \begin{subfigure}[h]{0.2\textwidth}
                \centering
\begin{tikzpicture}[]
\draw (15, 0) node[inner sep=0] { \includegraphics[scale = 0.22]{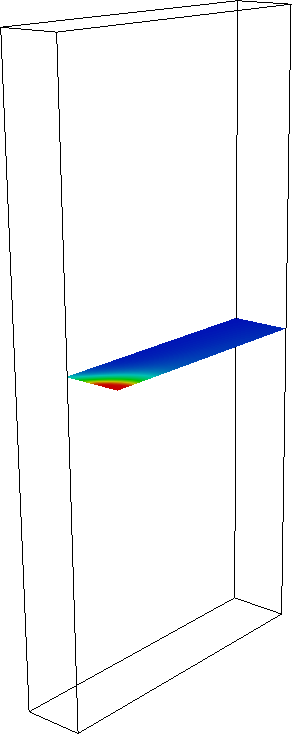}
\label{crack3d_thicka}};
\end{tikzpicture} 
                \caption{}
                \label{fig:3darrow}
        \end{subfigure}
        \begin{subfigure}[h]{0.35\textwidth}
                \centering
\begin{tikzpicture}[]
\draw (15, 0) node[inner sep=0] { \includegraphics[scale = 0.2]{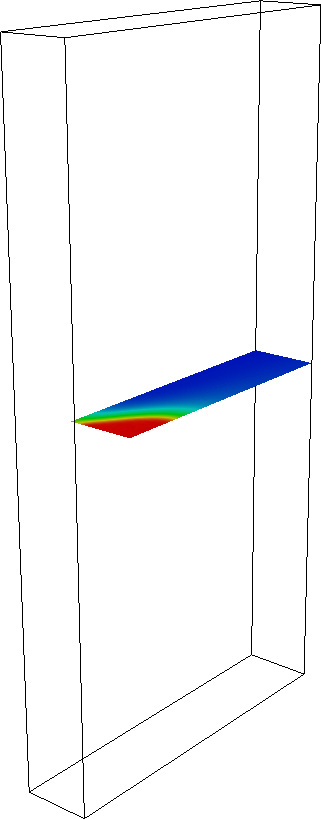}};
\end{tikzpicture}
                \caption{}
                \label{fig:3darrow1}
        \end{subfigure}
        \begin{subfigure}[h]{0.25\textwidth}
                \centering
\begin{tikzpicture}[]
\draw (15, 0) node[inner sep=0] { \includegraphics[scale = 0.2]{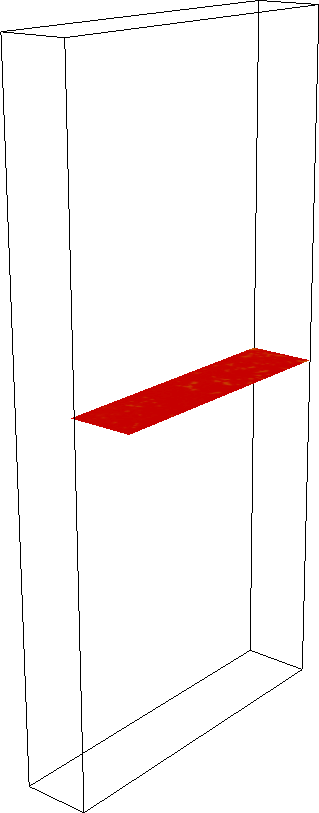}};
\draw (17.5, 0) node[inner sep=0]{ \includegraphics[scale=0.3]{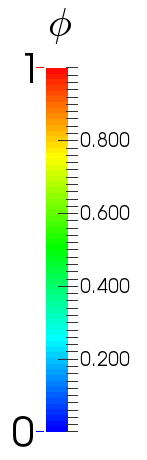}};
\end{tikzpicture}
                \caption{}
            \label{fig:3darrow2}
        \end{subfigure} 
        \caption{Crack extension for a graded plate with material properties varying through the thickness (FGM IV) at different loading stages, as given by remote displacements (a) $u = $ 5.137 $\times 10^{-3}$ mm, (b) $u = $ 5.165 $\times 10^{-3}$ mm, and (c) $u = $ 5.177 $\times 10^{-3}$ mm.}\label{fig:case5_crack3d_throughthickness}
\end{figure}

As shown in Fig. \ref{fig:3dforcevsDisp}, very different fracture responses can be obtained by placing the material gradient appropriately, relative to the crack surface. For example, in FGM III, where the crack is located on the edge with smaller $G_c$, cracking initiates early but substantial subcritical crack growth is observed, and final fracture occurs later than in the homogeneous specimen or the graded specimen where properties vary along $y$ (FGM I). This is due to the increasing fracture resistance encountered by the crack along the propagation path. On the other hand, FGM II exhibits a significant delay in crack initiation, relative to the other cases, but a sharp drop in the force at the onset of damage. Therefore, the present modelling framework can be employed to quantitatively design FGMs that exhibit a larger time to crack initiation or an enhanced crack growth resistance. In FGM IV, where properties change along the thickness, there is a spatially varying $G_c$ at the crack front. Thus, there will exist a weaker path for crack initiation, relative to the homogeneous case, which will trigger cracking at a lower load, see Fig. \ref{fig:3dforcevsDisp}. This is further explored by plotting the phase field parameter contours in Fig. \ref{fig:case5_crack3d_throughthickness}. As can be clearly observed, the crack initiates at the edge with smallest $G_c$, propagates along the $x$ and $z$ directions, and eventually fractures the specimen. 

\section{Conclusions}
\label{Sec:Concluding remarks}

A phase field formulation for fracture in functionally graded materials (FGMs) is presented. The framework has the capability of naturally capturing the gradient in fracture resistance inherent to FGMs via a spatially varying critical energy release rate $G_c$. A number of paradigmatic examples are addressed to emphasize the potential of the method in modelling the complex crack propagation paths that arise due to crack tip mode-mixity induced by the material property variation. Both two-dimensional and three-dimensional boundary value problems are solved. Unlike previous crack propagation studies on FGMs, the present formulation is proven capable of (i) predicting crack initiation from arbitrary nucleation sites, (ii) capturing crack deflection without re-meshing, and (iii) modelling unstable crack propagation in an implicit framework. The formulation can be readily extended to deal with other related material systems. For example, the extension to metal-based elastic-plastic FGMs is immediate in the absence of plastic-damage coupling, provided that plastic size effects are accounted for \cite{CS2018}. Similarly, the framework can easily accommodate other homogenization schemes, such as those proposed for functionally graded composites integrating carbon nanotubes \cite{Reinoso2016,Frikha2018}.\\

In addition, numerical predictions accurately reproduce experiments conducted in different types of functionally graded specimens, and under a wide range of material gradient profiles. The results obtained provide fundamental and quantitative insight into the role of the material property gradation on the crack propagation response. We identify combinations of volume fraction profiles and material gradient orientations that optimize crack growth resistance, laying the grounds for the design of fracture-resistant FGMs. These could be useful, for example, in components exposed to aggressive environments, where cracks and other stress raisers appear at the exposed surface.
 
\section{Acknowledgements}
\label{Acknowledge of funding}

E. Mart\'{\i}nez-Pa\~neda acknowledges financial support from the Royal Commission for the 1851 Exhibition through their Research Fellowship programme (RF496/2018).




\bibliographystyle{elsarticle-num}
\bibliography{library}

\end{document}